\documentclass[a4paper,12pt]{article}
\usepackage[english]{babel}
\usepackage[table, x11names]{xcolor}
\usepackage{url}
\usepackage{fullpage}

\usepackage[bottom]{footmisc}
\usepackage{graphicx}
\usepackage{color}
\usepackage{multirow}

\usepackage{amsthm}
\usepackage{amsmath}
\usepackage{amsfonts}
\usepackage{amssymb}
\usepackage{graphicx}
\usepackage{setspace}
\usepackage{natbib}
\usepackage{booktabs}
\usepackage{array}
\usepackage[skip=0pt,justification=raggedright,singlelinecheck=false,font=small,labelfont=bf,labelsep=endash]{caption}
\usepackage[FIGTOPCAP,nooneline]{subfigure}
\newcolumntype{K}[1]{>{\centering\let\newline\\\arraybackslash\hspace{0pt}}m{#1}}
\usepackage[colorlinks,breaklinks]{hyperref}
\hypersetup{
    colorlinks,%
    citecolor=black,%
    filecolor=black,%
    linkcolor=black,%
    urlcolor=black
}
\usepackage{lscape}
\usepackage{ctable}

\usepackage{array}

\usepackage{xcolor,colortbl}
\usepackage{eurosym}

\usepackage[toc,page]{appendix}

\usepackage{siunitx} 

\numberwithin{equation}{section}

\title{}
\author{
\\
\\
}

\begin{document}
	

\title{\Large{ \textbf{Shifting the yield curve for fixed-income\\and derivatives portfolios}\thanks{ The authors thank Antoine Bouveret, Cecilia Caglio, Alessandro Conciarelli, Maciej Grodzicki, Luc Laeven, Teodora Paligorova, participants to the ``2024 CEBRA Annual Meeting'' at Goethe University, Frankfurt, the ``ECONDAT Conference 2024'' at King's College London, and the 2024 workshop ``EMIR data analytics for research, financial stability and supervision'' at Banca d'Italia, Rome. The views in this paper are the authors' and do not necessarily reflect those of the Bank of Italy. Errors and omissions are the authors own responsibility.}
}}

\author{ \vspace{1cm} 
Michele Leonardo Bianchi\thanks{\textit{Financial Stability Directorate, Bank of Italy}, Rome, Italy~(\texttt{micheleleonardo.bianchi@bancaditalia.it})} 
\qquad 
Dario Ruzzi\thanks{\textit{Financial Stability Directorate, Bank of Italy}, Rome, Italy~(\texttt{dario.ruzzi@bancaditalia.it})}
\qquad 
Anatoli Segura\thanks{\textit{Financial Stability Directorate, Bank of Italy}, Rome, Italy~(\texttt{anatoli.seguravelez@bancaditalia.it})} \\
\today }
\date{\vspace{-0.1cm}}

\maketitle



\begin{abstract}
\noindent We use granular regulatory data on euro interest rate swap trades between January 2021 and June 2023 to assess whether derivative positions of Italian banks can offset losses on their debt securities holdings should interest rates rise unexpectedly. At the aggregate level of the banking system, we find that a 100-basis-point upward shift of the yield curve increases on average the value of swaps by 3.65\% of Common Equity Tier 1 (CET1), compensating in part for the losses of 2.64\% and 5.98\% of CET1 recorded on debt securities valued at fair value and amortised cost. Variation exists across institutions, with some bank swap positions playing an offsetting role and some exacerbating bond market exposures to interest rate risk. 
Nevertheless, we conclude that, on aggregate, Italian banks use swaps as hedging instruments to reduce their interest rate exposures, which improves their ability to cope with the recent tightening of monetary policy.
Finally, we draw on our swap pricing model to conduct an extensive data quality analysis of the transaction-level information available to authorities, and we show that the errors in fitting value changes over time are significantly lower compared to those in fitting the values themselves.
 
\end{abstract}

\vspace{0.5cm}


\vspace{0.5cm}

\newpage 
\renewcommand{\baselinestretch}{1.5} 
\setcounter{equation}{0}
\section{Introduction}
\label{sec:intro} 
Interest rate risk can be defined as the risk of losses on financial investments caused by adverse movements of yield curves. In this work we explore the interest rate risk of Italian banks' derivatives and debt securities holdings from a risk management perspective, that is we analyse the impact of potential yield curve shifts on the fair value of financial instruments, irrespective of any accounting or prudential rule. We rely on granular datasets on banks' portfolios available to Banca d'Italia. 
While bond data are taken from statistical and supervisory reports, information on derivatives is sourced from trade repository data collected under the European Market Infrastructure Regulation (EMIR).

A considerable portion of Italian banks' investments in securities -- which are commonly referred to as securities holdings -- involves fixed-rate bonds. The value of these debt securities declines when market interest rates rise, thus exposing banks to interest rate risk, which can be hedged  by taking positions in derivatives markets. 
In this paper we use the term ``hedge'' to refer to changes in value of bonds and derivatives of opposite sign following an interest rate shock. Nevertheless, it could be the case in practice that banks enter into derivatives to hedge not the bonds that we consider here but other assets such as retails loans, or even their liabilities. The purpose of derivative transactions is not reported in EMIR data and therefore hedging relationships cannot be established.

We analyse banks' interest rate exposures in the bond market with a methodology that complements the granular information on securities with market data from Refinitiv (see also \cite{bianchi2021var} and \cite{bds2022backtesting}). We then turn our attention to calculating, with a contract-by-contract full revaluation approach, banks' exposures in forward rate agreements and interest rate swaps referenced to EURIBOR, and overnight index swaps referencing EONIA and \euro STR. These contracts,  which we generically refer to as ``swaps'', are the most active EUR-denominated interest rate derivatives and represent a large share, in terms of gross notional amounts,  of Italian banks' derivative portfolios. 

Our paper makes two contributions. 
First, we contribute to the literature on interest rate risk in banking by measuring, with high accuracy and frequency, banks' capacity to hedge losses on security investments through exposures in interest rate swaps. Our monitoring framework delivers risk metrics that (i) hinge on functions overcoming the first-order risk exposure approximations usually found in the literature, and (ii) are estimated at a weekly frequency (potentially daily). To the best of our knowledge, this is the first study to investigate the interest rate risk of bank portfolios while accounting for these two features, which are extremely valuable from a risk management perspective. 
Second, we complement work on the quality of EMIR data, e.g.~\cite{esma-2023-transaction-data}, by examining the accuracy, consistency, and reliability of the swap contract values reported by market participants. Importantly, this is a valuable by-product of our framework that, by performing full revaluation (i.e.~pricing) of the swap contracts, delivers model-implied swap values to compare with those made available in reporting to authorities.   
The main objective of this paper is to enhance our understanding of the interest rate risk exposures of banks while providing insights on both risk assessment and data quality. This two-fold approach underscores the importance of reliable data for well-informed decision-making and effective monitoring within the dynamic and complex financial derivative landscape.

Our results spanning the period from January 2021 to June 2023 can be summarised as follows. First, on aggregate, swaps reduce Italian banks' exposures to interest rate risk arising from investments in fixed-rate securities, thus improving banks' ability to cope with the current juncture of monetary policy tightening. 
At the aggregate level of the banking system, we find that a 100-basis-point upward shift of the yield curve increases on average the value of swaps by 3.65\% of Common Equity Tier 1 (CET1), compensating in part for the losses of 2.64\% and 5.98\% of CET1 recorded on debt securities valued at fair value and amortised cost, respectively.  
Second, variation exists across institutions, with some bank swap positions playing an offsetting role and some exacerbating bond market exposures to interest rate risk. Nevertheless, the direction in which Italian banks in aggregate trade swaps, that is by benefiting from an increase in market interest rates, is consistent with hedging their business risk of borrowing short and lending long (maturity mismatch). This is also in line with evidence on swap trades provided for the European banking sector by \cite{hoffmann2019bears} and \cite{ECB_FSR_Nov22}, and for UK banks by \cite{khetan2023market}; in contrast \cite{McPhail2023_US_swaps} and \cite{jiang2023monetary} find that US banks have close to zero exposure in swaps after netting. 
Third, our data quality analysis of the granular derivatives information available to regulators unveils that it is easier to reconcile time changes in the reported swap values, rather than the values themselves, with the contract specific characteristics that are also being reported.

The remainder of the paper is organised as follows. Section \ref{sec:Literature} provides an overview of the literature on granular derivatives exposures. Section \ref{sec:Data} describes the EMIR and security holdings data. In Section \ref{sec:Methodology} we define the interest rate shock and we show how to compute swap and bond sensitivities. Section \ref{sec:empirical_findings} reports the main empirical findings on banks' interest rate risk exposures and on EMIR data quality checks. Section \ref{sec:Conclusions} concludes. 

\section{Related literature}
\label{sec:Literature}

To date, there have been relatively few empirical investigations of granular derivatives exposures, primarily due to limited data availability; see, e.g., the Annex of \cite{esma-2023-transaction-data}. At the European level, the implementation of EMIR in 2014 has required EU entities to report on a daily basis their derivative transactions to Trade Repositories (TRs), resulting in improved data availability. TR data are regulatory (non-public) contract-level data; for instance, in Europe, EMIR data are accessible only to European Securities and Markets Authority (ESMA), European Systemic Risk Board (ESRB), European Central Bank (ECB), and national competent authorities. Furthermore, they come in high volume and therefore a big data infrastructure is required for  their storage, manipulation, and analysis. The combination of confidentiality requirements and technical complexities has slowed down the development of a specific research avenue that leverages the granular derivatives exposures of market participants. Notably, scholarly articles investigating EMIR and similar regulatory data have only recently emerged in peer-reviewed academic journals.
Research focusing on granular derivatives exposures can be categorised into four main areas: (1) data description, encompassing data quality concerns; (2) deep dive into specific contract types, counterparties, or remarkable market occurrences; (3) systemic risk evaluation of margin practices and network structures; (4) risk sensitivity analysis. Our paper complements the first and last strands of literature.

Describing EMIR data is a multifaceted endeavour. After introducing the novel EMIR dataset and the procedures for its cleaning, \cite{abad2016shedding} offer statistical insights into the interest rate, foreign exchange (FX), and credit default swap (CDS) markets. Data cleaning methodologies are also discussed by \cite{perez-duarte-skrzypczynski-2019} and \cite{lenoci-letizia-2021}, while \cite{ascolese-et-al-2017}, \cite{van2017use}, and \cite{ullersma-van-lelyveld-2022} delve into the potential and obstacles of EMIR data for policymakers and regulators.

Studies centred on specific contract types include \cite{abbassi-brauning-2021}, who examine German banks' hedging strategies with FX forward contracts, and \cite{bahaj2023market}, who investigate the market of UK inflation swaps. Then there are several analyses of CDS contracts, encompassing the works of \cite{kenny-et-al-2016}, \cite{levels-et-al-2018}, \cite{abraham-2020}, \cite{pedro-2020}, \cite{bianchi-2021}, \cite{mazzacurati-et-al-2021} and \cite{bilan2023bank}. Within the realm of CDS, single-name contracts stand out due to their standardised information and readily identifiable risk direction. 

Studies centred on specific counterparties include \cite{bellia-et-al-2019}, who scrutinise the clearing choices of counterparts engaged in CDS linked to sovereign debt of Italy, France, and Germany, \cite{bias-et-al-2020} who investigate the use of derivatives by equity mutual funds, while \cite{kenny-et-al-2016} and \cite{fiedor-killeen-2021} focus on securitization special purpose vehicles. 

Researchers have also used granular derivatives data to investigate specific market events. \cite{cielinska-et-al-2017} and \cite{joseph-vasios-2022} delve into the repercussions of the Swiss National Bank's decision to eliminate the floor of 1.20 Swiss francs per euro on the FX derivatives market in January 2015. \cite{schroeder-et-al-2020} analyse the underlying factors behind the Pound Sterling vs US Dollar flash crash in October 2016, including the role of various market participants in the liquidity disruption during that event. \cite{bouveret-haferkorn-2022} discuss and show how to trace the significant surge in concentrated exposures that the family office Archegos undertook in early 2021.

Several papers assess initial and variation margins, with a focus on systemic risk and risk management practices of central counterparties. This research area includes the works of \cite{cominetta-et-al-2019}, \cite{esrb-2020-procyclicality}, and \cite{grothe-et-al-2021}, as well as \cite{ghio2023derivative}. \cite{derrico-roukny-2021} explore the mechanics of derivative portfolio compression in terms of both feasibility and efficiency. \cite{jukonis2022impact} study the liquidity risk of euro area investment funds, specifically stemming from the demands of variation margin calls on their derivative holdings. 

A number of studies feed EMIR data into network approaches to monitor and assess systemic risk arising from derivative markets. \cite{zema2023uncovering} uncovers the collateral network structure of non-centrally cleared derivatives, while \cite{fiedor-et-al-2017} provide a comprehensive analysis of the EU centrally cleared interest rate derivative market. \cite{bardoscia-et-al-2019} create separate networks for interest rate, credit, and FX derivatives markets, which they then amalgamate into a multiplex network to gauge the systemic vulnerability of individual institutions with centrality metrics tailored for multiplex networks. \cite{rosati-vacirca-2020} reconstruct the relationship network in the centrally cleared derivatives market among clearing members, CCPs, and clients. They analyse the network's topology using various centrality metrics, identifying pivotal nodes and evaluating their interconnectivity across market segments. Using derivative margins and cash buffers data, \cite{Bardoscia2021} develop a model for estimating liquidity shortfalls during stress scenarios. \cite{kahros-et-al-2021} present a descriptive analysis of client clearing in OTC and exchange-traded derivative markets across various products.

A newer research stream has recently emerged, primarily focused on risk assessment. \cite{jukonis-2022} introduces a model to estimate market risk for equity derivative exposures under adverse scenarios. \cite{khetan2023market} explore EMIR data to provide the first large scale empirical evidence on risk sharing in the interest rate swap market. Their study reveals that pension and insurance sectors often offset risk from banks and corporations. However, the segmentation of demand across different maturities and the behavior of hedge funds complicate risk offsetting, impacting dealer positions and swap spreads. Additionally, long-term pension fund trades, which are less frequently centrally cleared, introduce counterparty credit risk. Using comprehensive data spanning UK government bonds, interest rate swaps, options, and futures markets, \cite{pinter2023hedging} assess interest rate risk for non-bank financial intermediaries. They find that these intermediaries tend to employ derivatives not to hedge but rather to amplify their interest rate risk exposures in the bond market. They also document high market concentration that could impair the transmission of monetary policy and exacerbate the effects of idiosyncratic shocks. \cite{McPhail2023_US_swaps} show that interest rate swaps are not economically significant for hedging the risk of long-term assets of US banks. The notional amount of swaps held by these banks is substantial, however, after accounting for offsetting swap positions, the typical bank has minimal exposure to interest rate risk, with a 100-basis point rate increase affecting the value of its swaps by just 0.1\% of its equity. 

\cite{hoffmann2019bears} investigate how interest rate risk is distributed among European banks sector using supervisory data including EMIR. To measure interest rate risk, they evaluate the impact on the fair value of balance sheet items in response to a shift of the interest rate curve and show that banks use interest rate swaps to hedge fair value risk, but not income risk. However, their approach is different from ours. They analyse the entire bank balance sheet and perform an econometric analysis to explain banks' exposure to interest rate risk. Their study is focused on repricing cash flows pertaining to assets and liabilities in the banking book and on EURIBOR and EONIA interest rate swaps as of December 31, 2015. Instead, our work has a risk management perspective: the estimates are conducted at a weekly frequency (and can be easily extended to daily) because we are interested in measuring and studying the dynamics of the interest rate risk exposures in all bonds (in both banking and trading book) and swaps. Additionally, we exploit all available granular information at contract level, for both bonds and derivatives, and perform a revaluation to estimate the risk of bank portfolios.

\section{Data}
\label{sec:Data}

In this paper, we combine two regulatory and highly granular datasets. First, we use transaction-by-transaction EMIR data on derivatives to retrieve interest rate swap trades reported by Italian banks. Second, for each bank with swap positions, we use security-by-security holdings data to define their investments in fixed-rate bonds. We provide detailed descriptions on these datasets and how the sample is constructed below.

\subsection{EMIR}
\label{sec:EMIR_Data}

We obtain data on interest rate swap contracts in accordance with European Market Infrastructure Regulation (EMIR).\footnote{See Regulation EU/2012/648.}  According to this regulation, transaction-by-transaction derivatives data are reported on a daily basis by entities resident in the EU since February 2014 and collected through trade repositories, which, in turn, make these data available to authorities. Stored in over 100 data fields, the collected information includes details of each individual derivative transaction such as the identity of the counterparties involved, the type of derivative, the contract value, its maturity and notional amount outstanding, the underlying security, the execution and clearing venues, and, if any, the collateral (margin) paid and received. 

In this paper, we work with EMIR data accessible to Banca d'Italia, that is data as of December 29, 2020, for entities and markets falling within the jurisdiction and financial stability mandate of Banca d'Italia.\footnote{\cite{GdL_EMIR_2023} provides an exhaustive description of the full EMIR data and the portion of it that is accessible to Banca d'Italia. The authors also highlight data quality and practical issues faced when collecting and using EMIR data, and define the framework used in this paper to move from raw to clean data suitable for the analyses.} 
We focus on a subsample restricted to the euro-denominated interest rate derivatives of Italian banks consolidated at group level.\footnote{Consolidation is based on Banca d'Italia's regulatory group structure information in cases where the group parent is a bank that resides in Italy, whereas for all other groups it is based on information reported by the Global Legal Entity Identifier Foundation (GLEIF) established by the Financial Stability Board (FSB) in June 2014, to foster the widespread use of identifiers for legal persons.} Although EMIR data is recorded daily with about 30 million records available to us for each trading day, we choose, for practical reasons, to conduct the analysis at weekly frequency by sampling swap trades outstanding on the Wednesday of each week from January 2021 through June 2023. We use ``trade state'' data, which contains all pending trades at the end of a given day. 

Reporting under EMIR is dual-sided, i.e.~both counterparties to derivative contracts are required to report if they fall under the scope of EMIR. Therefore, for some trades, our dataset contains two reports of the same trade, one by each counterparty. We identify these duplicate trades using the ``trade id'' field in EMIR, and we sample only one report per trade to avoid double counting of a single trade. For each entity of interest, i.e.~a bank, whenever two reports are available we keep the report submitted by that entity.

We now describe in detail how we select the contracts analysed in the empirical part and we show some statistics related to the Italian banks' derivative portfolios. We focus on forward rate agreements (FRA) and interest rate swaps (IRS) written on the euro interbank offered rate (EURIBOR), and overnight index swaps (OIS) written on the euro overnight index average rate (EONIA) and the euro short-term rate (\euro STR).\footnote{EONIA was discontinued on 3 January 2022 but OIS contracts referenced to this rate continue to exist until their expiration. We treat these contracts as though they were OIS referencing \euro STR.}
Among these the EURIBOR swaps are the most traded and liquid
derivatives used to hedge interest rate risk for euro-denominated exposures \citep{ECB_FSR_Nov22}.
We deal with single-currency fixed-for-floating contracts, both spot-starting and forward-starting trades.\footnote{It should be noted that in a forward-starting swap, the exchange of payments does not begin until an agreed future date, $T_1$, after which they continue to the maturity date, $T_2$. Thus, a long (i.e.~pay fixed-rate) position in a forward-starting swap is equivalent to a long position in a spot-starting swap maturing at $T_2$ and a short position in a spot-starting swap maturing at $T_1$.}
We use the term ``swaps'' when referring not only to IRS and OIS but also to FRA, the reason being that FRA are single-period IRS for forward settlement. 

We identify the candidate swaps in EMIR as trades where: (i) the ``asset class'' field is equal to ``IR'' (for ``interest rate''), (ii) the ``contract type'' field is equal to ``SW'' (for ``swap'') or ``FR'' (for ``forward rate agreement''), (iii) either the ``floating rate of leg 1'' field or the ``floating rate of leg 2'' field is not blank, and whichever field is populated contains the term ``euri'', ``str'', ``eona'', or ``eonia'' (after lowercasing the text), (iv) either the ``fixed-rate of leg 1'' or the ``fixed-rate of leg 2'' field is not blank, and (v) only one of the ``notional currency 1'' and ``notional currency 2'' fields is not blank, or they both contain the same value. Finally, we identify forward-starting IRS and OIS contracts as those swaps where the ``effective date'' field is past the ``reference date'' of the reporting. We gather from EMIR the contractual features of every position on these candidate swaps at each bank in the sample. 

In a fixed-for-floating swap, one party agrees to pay a fixed-rate and to receive a floating rate on some notional amount for a fixed term, while the other party agrees to pay that floating rate and to receive that fixed-rate on the same notional amount for the same term. Here and thereafter, we adopt the following convention: the counterparty that pays fixed is the buyer of the swap (``long'' position), while the counterparty the receives fixed is the seller of the swap (``short'' position).\footnote{We use the ``counterparty side'' field reported in EMIR, which is populated with either ``B'' for buy or ``S'' for sell, to distinguish between long and short positions. In Regulation EU/2013/148, the reporting rules of EMIR state, {\it ``In the case of an interest rate derivative contract, the buy side will represent the payer of leg 1 and the sell side will be the payer of leg 2''}, where each leg can only be populated with either the fixed or the floating rate. More recently, Regulation EU/2017/105 has amended the reporting rules by stating, {\it ``In the case of swaps related to interest rates or inflation indices, the counterparty paying the fixed-rate shall be identified as the buyer and the counterparty receiving the fixed-rate shall be identified as the seller''}. Taking into account this change of rules, we identify whether the reporting counterparty of a swap in our data is paying the fixed or floating rate by repricing the contract under both interpretations of the ``counterparty side'' field. We always choose the interpretation corresponding to the more recent regulation unless the repricing according to it, and not the other regulation, delivers a swap contract value with opposite sign from that of the value reported in EMIR (even after rescaling the former to have the same time series average as the latter). What we observe in practice is that, with very few exceptions corresponding to 5\% of the sampled trades, entities submit their reports in accordance with the more recent regulation.} 
A swap's interest rate payments are exchanged regularly throughout the life of the contract, e.g.~twice a year.\footnote{In a FRA, counterparties exchange on a future date (settlement date) a single interest payment at a fixed rate for a payment at the then-prevailing value of a floating rate applicable to an investment starting on that date and lasting until maturity date of the FRA.} The fixed-rate, which is called the swap rate, is determined at the time of the trade and is normally set such that the value of the swap at initiation is zero, or in other words, the payment of an upfront amount by one counterparty to the other is not required to enter into the swap. The floating rate, which is usually a compounded overnight rate or an interbank offered rate, is in our case the \euro STR or an EURIBOR of specified maturity, whose future realisations determine the floating rate payments that will be exchanged for the fixed-rate payments. We filter the trades referencing EURIBOR to keep only contracts where the floating rate has a maturity that is: (i) either 1, 3, 6, or 12 months, (ii) exactly equal to the number of months between two consecutive rate resets (that is when the rate is measured to determine the payments), and (iii) also equal to the number of months between two consecutive payment dates. 
As for the fixed-rate, the data manipulation and cleaning we perform consist of first using the ``price/rate'' field whose value, when provided and valid, expresses the swap rate in percent values. When a valid value is not available, in the second step we resort to the ``fixed-rate of leg 1'' and ``fixed-rate of leg 2'' fields. We compare the reported fixed-rate of a given contract with the par swap rate (sourced from Refinitiv) that prevailed on the market when obligations under that contract come into effect (``effective date'' field), for a term equal to the original maturity of the contract. Such a comparison allows us to understand when the reported fixed-rate is not yet expressed in percentage and when is an outlier (below -2.5\% and above 25\%) that needs replacing with the market par swap rate.

After applying the aforementioned data-filtering criteria and discarding trades whose valuation is dated more than one week prior to the reporting date, our sample includes 218,717 unique swap contracts traded by 54 banks, for a total of 14,664,747 swap-week observations covering the sample period from January 2021 to June 2023.\footnote{We have excluded 4 intermediaries for which we do observe swap positions in EMIR but we do not have data on euro fixed-rate bonds or CET1 capital.} 

\begin{figure}[h!] 
\caption{Swaps traded by Italian banks}
\label{fig:IRS_notional}
\vspace{0.1cm}
\subfigure[Gross notional outstanding]{
	\includegraphics[width=7.7cm, height=6.0cm]{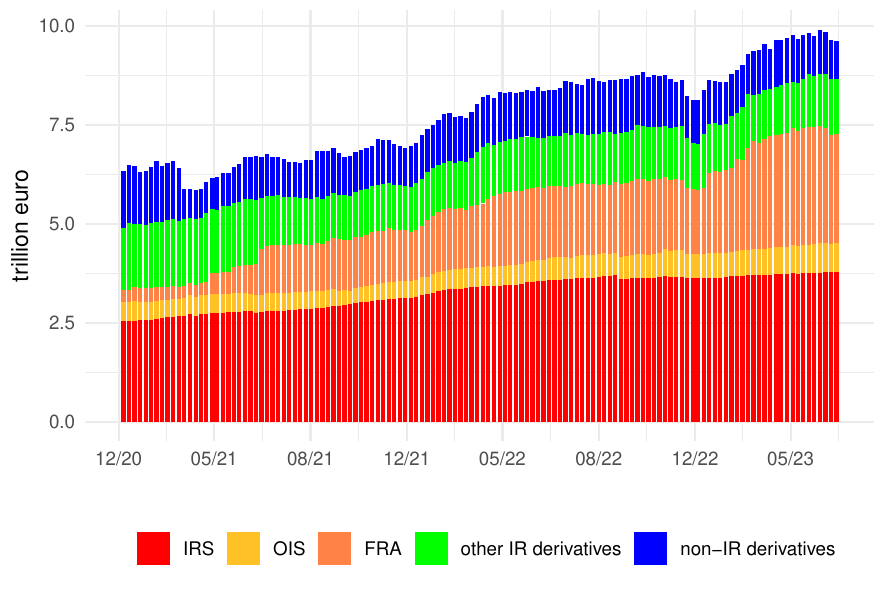}  
}~~ 
\subfigure[Net notional positions by maturity]{
	\includegraphics[width=7.7cm, height=6.0cm]{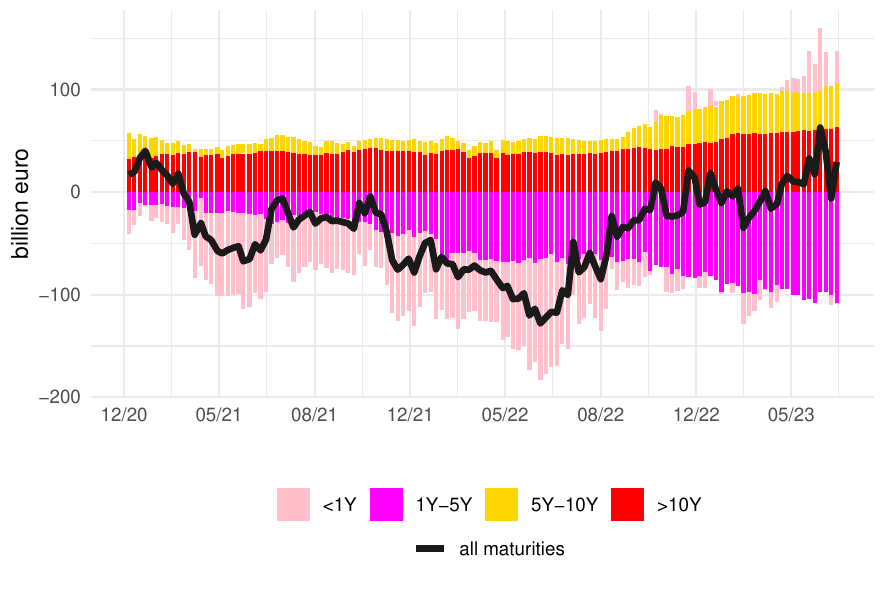} 
}
{\footnotesize 
\begin{spacing}{1}
	Notes: Panel (a) displays the aggregate gross notional (sum of long and short notional) exposures of Italian banks in EURIBOR swaps (red), EONIA and \euro STR overnight index swaps (yellow), EURIBOR forward rate agreements (orange), other interest rate derivatives (green), and non-interest-rate derivatives (blue); values are in EUR trillion. Panel (b) displays the aggregate net notional (difference between long and short notional) exposures of Italian banks in swaps (IRS, OIS and FRA) when considered as a whole (solid black line) and when aggregated in four time-to-maturity buckets; values are in EUR billion with positive numbers denoting net long (pay-fixed) positions and negative numbers denoting net short (receive-fixed) positions. Data are weekly and sampled each Wednesday from January 2021 to June 2023. 
\end{spacing}}	
\end{figure}

As shown in Panel (a) of Figure \ref{fig:IRS_notional}, the selected swap positions, which include euro-denominated IRS, OIS and FRA, represent a sizeable share of derivative portfolios held by Italian banks. By the end of our study period, they account for about 76\% and 84\% of the gross notional amount (i.e. the sum of long and short notional amounts) of, respectively, all and interest-rate-only banks' derivative exposures. Against a background of monetary policy normalisation and tightening, bank activity in these swaps has intensified, with the gross notional outstanding more than doubling since the start of 2021 to \euro 7.3 trillion. This figure is made up of \euro 3.8 trillion gross notional from IRS, \euro 2.8 trillion from FRA, and \euro 736 billion from OIS.
Notional amounts, which only define the size of the interest payments exchanged, cannot be interpreted as measuring the interest rate exposure of swap positions, which depends on other contract characteristics, such as the rate and maturity of the swaps \citep{baker2021risk}. For instance, FRA, which tend to have a maturity of 1 year, are much less exposed to interest rate risk than the longer-dated IRS and OIS.  
Furthermore, the large notional amounts just described do not take into account that long and short positions offset each other, thus effectively reducing the interest rate risk exposure in swaps. 
This is even more critical for banks, which, because of their role as market makers and clearing members of central clearing counterparties (CCPs), often intermediate trades by entering into a contract with a client and an identical contract of opposite direction with the CCP. 
The solid black line in Panel (b) of Figure \ref{fig:IRS_notional} shows that, as of June 2023, the \euro 7.3 trillion gross notional amount of swaps in the Italian banking system falls to \euro 29 billion after netting.\footnote{This netting approach continues to suffer from not considering contract characteristics other than notional amount. Our preferred metric of interest rate exposure, which looks at the change in the fair value of an asset in response to a change in interest rates, will be discussed in the following sections.} 
The net value is positive, implying that the notional of long (pay-fixed) positions exceed that of short (receive-fixed) ones. 
On aggregate and in terms of traded notional, banks were strongly net receivers of fixed-rate payments until immediately before the first rate hike by ECB in July 2022. Since then, net short positions have significantly decreased (i.e.~the solid black line becomes less negative and turns positive) turning into net long positions as well. This evidence is consistent with the expectation, in the months after the monetary policy tightening started, of higher interest rates in the future. 
In Panel (b) we also report the net notional positions on swaps aggregated in four maturity buckets. The 1Y-5Y and $>$10Y maturities stand out with large net notional amounts, the former with a net short position of \euro 109 billion at the end of the sample, and the latter with a net long position of \euro 64 billion.

\subsection{Securities holdings}
\label{sec:SH_Data}

After having described derivatives data, we briefly illustrate the granular data on bank securities portfolios collected by Banca d'Italia. Starting from 2008, monthly security-by-security holding data at an individual level are collected for each bank or subsidiary located in Italy.  These portfolios are aggregated at a consolidated level by removing intragroup exposures. 

The data are collected at ISIN-by-ISIN level.\footnote{ Recall that the International Securities Identification Number (ISIN) is the internationally recognized code for the identification of financial instruments in the markets and in transactions. It is based on the ISO 6166 standard.} It should be noted that these are statistical data and all securities are evaluated at fair value, irrespectively of the accounting portfolio. Furthermore, for banks, there are no information on prudential portfolios and we do not know if a security belongs to the trading book or to the banking book.\footnote{From the end of 2018 certain banking groups (e.g.~for Italy, all Italian significant institutions) must provide, on a quarterly basis, more detailed statistics on their securities holdings in respect of so-called Securities Holdings Statistics Group (SHSG) data directly to the \cite{shsg2018}.} 

Additionally, the Banca d'Italia maintains an electronic archive, named Securities Database, containing the details of each ISIN that banking and financial intermediaries and other companies report. The characteristics needed to define the securities risk factors can be directly extracted from this database (i.e.~security type, issuer sector, issuer country or region, residual maturity, coupon type and currency). This allows us to explore and analyse in an efficient and automated way bank portfolio risks.

The dataset on which our empirical study is based includes end-of-month portfolio allocations from December 2020 to June 2023. Since our sensitivity analysis is conducted at weekly frequency starting from the first week of January 2021, we rely on the assumption that portfolio allocations are held constant until the next month-end date.
As already observed, we consider the sample of intermediaries having both securities and derivatives exposures on the risk factor analysed in this study  (i.e., the euro interest rate curve) for a total of 54 intermediaries (see Section \ref{sec:EMIR_Data}).


\begin{figure}[h!] 
\caption{Securities holdings of Italian banks}
\label{fig:Portfolios}
\begin{center}
\vspace{0.1cm}
\subfigure[Portfolio composition]{
	\includegraphics[width=7.7cm, height=6.0cm]{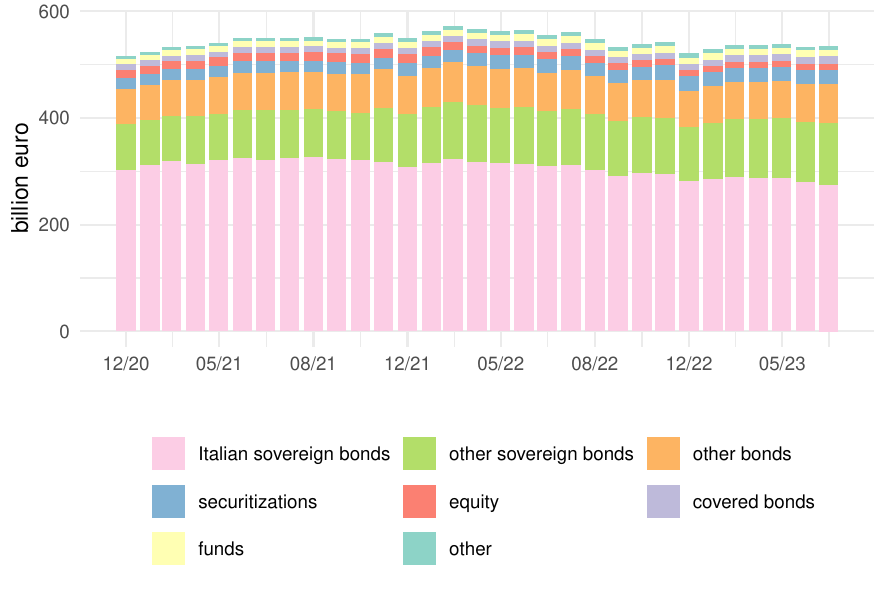}  
}~~ 
\subfigure[Euro fixed-rate bonds]{
	\includegraphics[width=7.7cm, height=6.0cm]{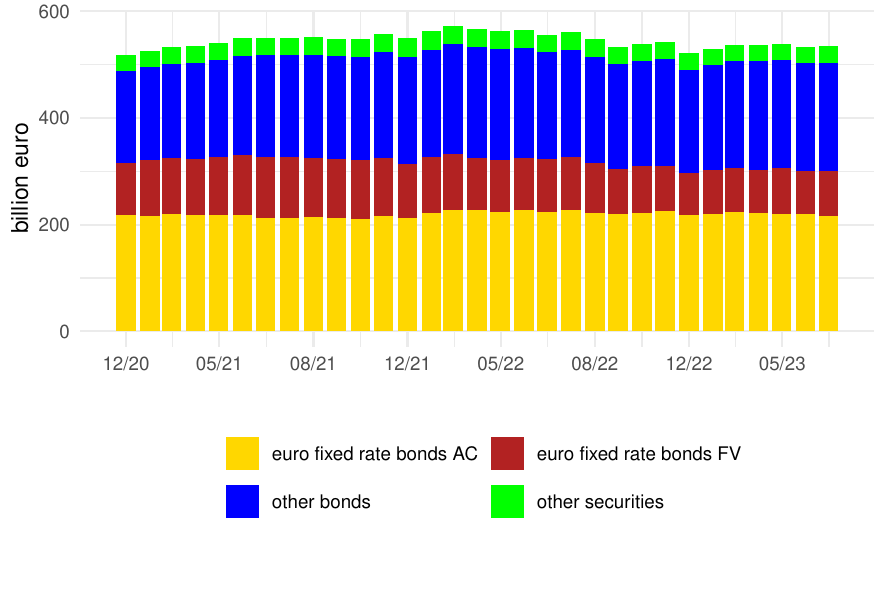} 
}
\subfigure[Residual maturity]{
	\includegraphics[width=7.7cm, height=6.0cm]{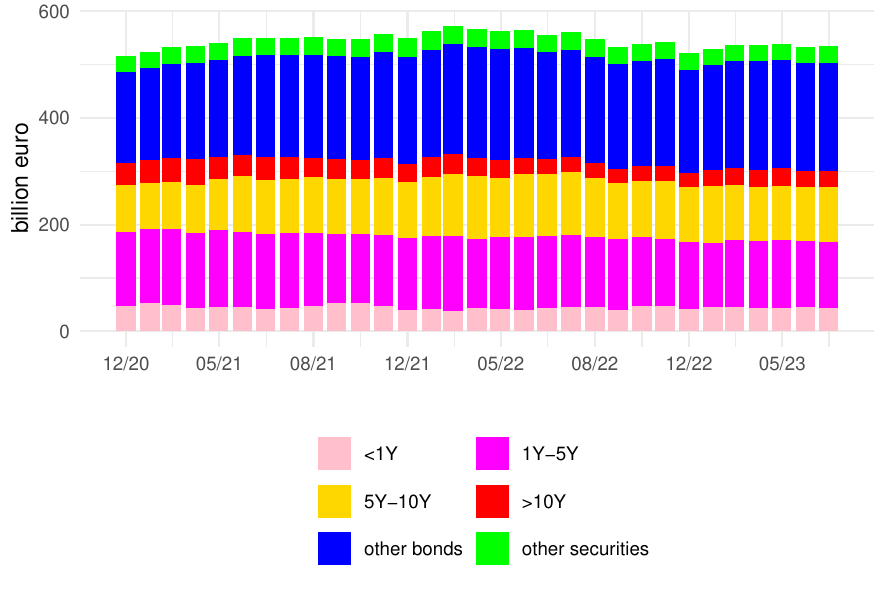} 
}
\end{center}  
{\footnotesize 
\begin{spacing}{1}
	Notes: Panel (a) displays, for the sample of 54 intermediaries, the aggregate portfolio composition by security type. Panel (b) shows the same portfolios by focusing on euro fixed-rate bonds valued at amortised cost (AC) and at fair value (FV). Panel (c) displays, for the same sample, the aggregate residual maturities of the bonds considered in this study. Other bonds and other securities do not depend on the risk factors analysed in our empirical study. All data are monthly from December 2020 to June 2023. 
\end{spacing}}	
\end{figure}

In Figure \ref{fig:Portfolios}, for the sample of 54 intermediaries considered in this study,  we show the aggregate portfolio composition and the residual maturities of the assets in the portfolios. The maturity is shown only for euro denominated fixed-rate bonds, that is for those bonds directly affected by a shift of the euro interest rate curve. The assets labelled as {\it other bonds} do not depend on the risk factors analysed in our empirical study and for this reasons we do not show their residual maturity. The assets labelled as {\it other securities} do not have a specified maturity. Intragroup transactions and participations are excluded. 

From December 2020 to June 2023, the fair value of the aggregate securities portfolio of the Italian intermediaries in the sample was on average \euro 545 billion, with a peak of \euro 573 in February 2022 (Panel (a) of Figure \ref{fig:Portfolios}). For the sample analysed in this study, Italian sovereign bonds represent a large portion of the investments (56\%). Bonds issued by foreign sovereign or international bodies represent around 18\%. The exposure to equities and funds is limited, less than 5\%, while the exposure to corporate and financial bonds is remarkable (slightly less than 13\%). Securitizations and covered bonds are around 6\% of the aggregate portfolio fair value.

As shown in Panel (b) of Figure \ref{fig:Portfolios}, euro fixed-rate bond represent, on average, more than half (58\%) of the securities holdings of the banks in the sample: bonds valued at amortised cost (AC) and at fair value (FV) represent 40\% and 18\% of the total portfolio, respectively.

In Panel (c) we display the residual maturity of the bonds in the portfolio of the analysed intermediaries, the 1Y-5Y and 5Y-10Y maturities stand out with large amounts, the former with an average amount over the analysed period of 24\% of the total, and the latter of 19\%. Recall that the assets labelled as {\it other bonds} ({\it securities}), which amount, on average, to almost 36\% (6\%) of the total portfolio value, are not analysed in this study. 

Our sensitivity analysis includes bonds valued at both AC and FV. We consider bonds present not only in the banking book but also in the trading one because we aim to understand the interest rate risk of banks from a risk management or financial stability viewpoint rather than from a prudential one.

\section{Methodology}
\label{sec:Methodology}

This section describes the framework used to measure the interest rate risk from banks' portfolio in the euro interest rate swap and fixed-rate bond markets. We first define the underlying risk factor driving the asset prices and we assume a stress scenario where it takes on an extreme value. We then explain how to compute per-contract valuation changes in a bank's swap and bond portfolios after the shock. These changes represent the main inputs of the sensitivity analysis, in which we compare the two portfolio impacts to determine whether, in part or in total, they offset each other.







Owing to the asset class that is the focus of this study, the market shock that would trigger significant asset price fluctuations is a movement in interest rates. More specifically, the single underlying risk factor that we examine is the euro risk-free spot ``yield curve'' used to discount swap and bond expected future cash flows back to present day. As a shock we apply a 100 basis point parallel upward shift to the yield curve. We assume that the shock, which we implement as an instantaneous shock, passes through in full both to the forward rates that define the swaps' floating rate payments, and to the bonds' yield-to-maturity, implying that bond issuers' credit risk remains unchanged.
To put the 100 basis point shock into perspective, this value is more than three times larger than any single-day moves of 12-month EURIBOR rates in history, including the 29 basis point change in June 2008 at the onset of the global financial crisis and the 15 basis point change in July 2022 with the normalisation of the monetary policy by ECB.
Although extremely unlikely within a day, a 100 basis point move in rates may be observed over longer time horizons like the two-month periods in the second half of 2022 in our sample.
In Appendix \ref{sec:bootstrap_rates} we provide detailed information on how to bootstrap the riskless spot (i.e.~zero-coupon) curve that provides discount rates of swap and bond pricing and is the object of the shock in our sensitivity analysis.\footnote{In the case of derivatives, including swaps, discount rates and risk-free rates are the same, whereas the discount rates of bonds have an extra component (spread) that reflects the credit risk of the issuer.}

Having defined the underlying risk factor, we are now able to elucidate swap and bond risk sensitivities.
We compute banks' interest rate risk exposure in swaps (i.e.~swap sensitivity) as the change in value of their swap portfolios after the shock. We do this by repricing each individual swap contract before and after the shift in the yield curve, and then taking the difference between the pre- and post-shock values. Unlike the approximations that will be discussed later for bonds, this is a full revaluation approach that delivers changes in market values of contracts computed using comprehensive pricing formulae. We favour this method because, from a broader monitoring perspective, we are also interested in checking the data quality of the contract value reported in EMIR by comparing it against the price returned by our model.
This analysis would not be possible using the first-order approximations for swap sensitivities proposed by \cite{Bardoscia2021}, among others. 
The comprehensive pricing formulae of swaps are provided in Appendix \ref{sec:pricing_formulae} and are based on the interpretation of a contract as a long/short combination of a bond paying the fixed-rate on the swap and a floating-rate bond paying the money market reference rate. This means that a short (i.e.~receive-fixed) swap position can be valued as the price of the fixed-rate bond minus the price of the floating-rate bond, that is
\begin{equation}
	\label{eq:swap_value}
	V^{swap} = B^{fix} - B^{fl} ~ .
\end{equation}

To evaluate the impact of the yield curve shift on the euro-denominated fixed-rate bonds, we consider a partial revaluation approach based on a second-order approximation (delta-gamma) that relies on bond modified duration and convexity. Unlike the full revaluation approach adopted for swaps, we choose to keep the bond pricing framework as simple as possible. This is because the approximation error is small for this type of assets and the input data needed for the estimation can be directly obtained from Refinitiv without a comprehensive understanding of all characteristics of the bonds. 
Full details on the procedure underlying bond sensitivity are in Appendix \ref{sec:pricing_formulae}.

\section{Empirical findings}\label{sec:empirical_findings}

In this section we present our empirical results. We first investigate in Section \ref{sec:swap_pricing_fit} the accuracy of the swap pricing model (\ref{eq:swap_value}) relative to the values reported in EMIR. We consider the pricing error not only in the contract values but also in their time-series changes, as a small error in the latter is what matters most in sensitivity analysis. In Section \ref{sec:sensitivity} we discuss the estimates of banks' interest rate risk exposures, which allow to determine if and how effectively banks use swaps to hedge risk of their fixed-rate bonds.

\subsection{Swap pricing fit}
\label{sec:swap_pricing_fit}

\begin{figure}[!htb]
\caption{Goodness of fit of swap contract values}  
\label{fig:CV_fit_scatter}
\vspace{-0.5cm}
\begin{center}
	\includegraphics[width=11cm, height=9cm]{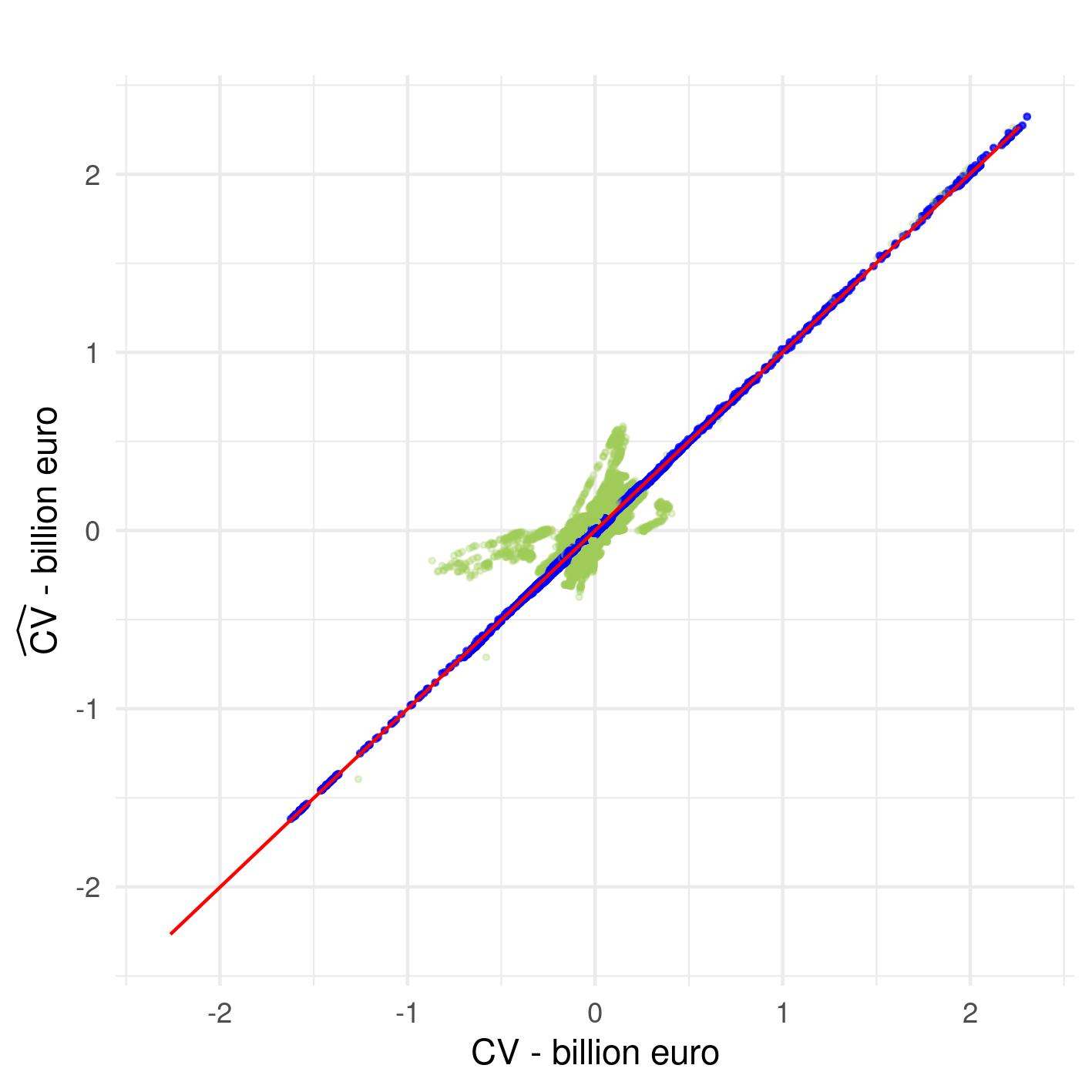} 
\end{center}
\vspace{-0.5cm}
{\footnotesize 
\begin{spacing}{1}
	Notes: The figure displays the fit between the swap contract values reported in EMIR, $CV$, and those implied by the pricing model (\ref{eq:swap_value}), $\widehat{CV}$. Each dot represents a swap-week observation in the sample (14,664,747 in total). Green dots denote observations (43,867 in total) with absolute pricing error greater or equal to \euro 25 million. 
	The regression line, in red, is estimated to have slope of 1.001 with robust method.
\end{spacing}}
\end{figure}

We start the assessment of the swap pricing model (\ref{eq:swap_value}) by comparing the model-implied contract values, which we denote by $\widehat{CV}$, to the values reported in EMIR, which we denote by $CV$. The goodness of the model in fitting the swap contract values is shown in Figure \ref{fig:CV_fit_scatter}, where each dot represents a swap-week observation in our dataset. 

\begin{figure}[!htb] 
	\caption{Swap contract values and model fitting errors}
	\label{fig:error_CV}
	\vspace{-0.2cm}
	\begin{center}
		\subfigure[~~~~~~~~~~~~~~~~~~~~~~~~~~~~~~~~~~~~~~~~~~~~~~~~~~~~~~~~~~~~~~~$|CV|$]{
			
			\includegraphics[width=16cm, height=3.8cm]{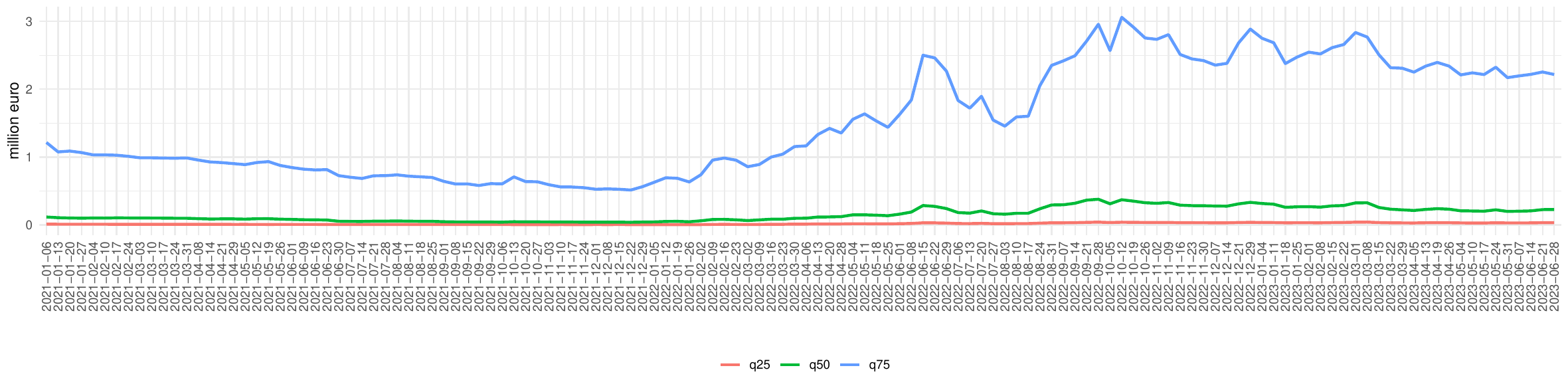}  
		}
		\\ 
		\subfigure[~~~~~~~~~~~~~~~~~~~~~~~~~~~~~~~~~~~~~~~~~~~~~~~~~~~~~~~~~~~~$|CV - \widehat{CV}|$]{
			\includegraphics[width=16cm, height=3.8cm]{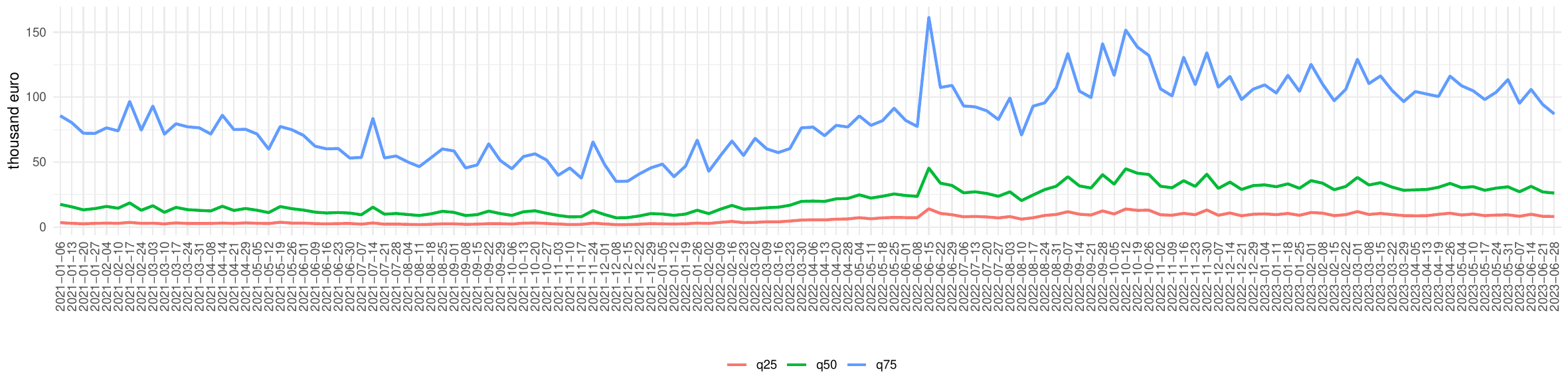} 
		}
	\end{center}  
	\vspace{-0.5cm}
	{\footnotesize 
		\begin{spacing}{1}
			Notes: Panel (a) displays the time trend of the 25th, 50th, and 75th percentiles of the distribution of the absolute contract values reported in EMIR. Panel (b) displays the time trend of the percentiles of the absolute difference between the contract values reported in EMIR and those implied by the model (\ref{eq:swap_value}). 
	\end{spacing}}	
\end{figure}

\noindent Overall we find a good fit between the data and the model, as indicated by the almost unitary slope (1.001) of the estimated line obtained with robust regression (the M-estimation method). At the same time, we observe a relatively large amount of dots departing from the regression line due to substantial pricing errors (we highlight in green the observations with absolute pricing error greater or equal to \euro 25 million). Nevertheless, 
it is reassuring that these dots are mostly located in the first and third quadrants, thus indicating a good correspondence of sign between the model-implied and reported contract values. It is important to note that pricing errors can be the result not only of erroneous inputs fed into the model (e.g.~a fixed-rate measured on the wrong scale or an unreported spread paid on the floating leg) but also of erroneously reported contract values against which we compare the model estimates. Furthermore, some discrepancies are possibly due to adjustments made to the fair value of  derivatives contracts that the trading entities include in their pricing framework to take into account funding, credit risk and regulatory capital costs. These potential trade-by-trade valuation adjustments  are  not accessible to us.
To help visualise the magnitude of the pricing errors, we show the time trend of the interquartile range -- that is the 25th, 50th, and 75th percentiles of the distribution -- of their absolute values in Panel (b) of Figure \ref{fig:error_CV}. Calculated across all dates, the median absolute error is \euro 21,182. At the end of the sample, the value is \euro 26,130, considerably higher than the \euro 17,438 observed at the start. The increase is at least in part a result of larger contract values reported in EMIR since 2022, when interest rates began to rise steadily. In fact, as it can be seen from Panel (a) of Figure \ref{fig:error_CV}, the median absolute contract value in EMIR has almost doubled since the start of year 2021 to \euro 225,737. 

%
%

\begin{figure}[!ht]
	\caption{Goodness of fit of weekly changes in swap contract values}  
	\label{fig:DeltaCV_fit_scatter}
	\vspace{-0.5cm}
	\begin{center}
		\includegraphics[width=11.0cm, height=9cm]{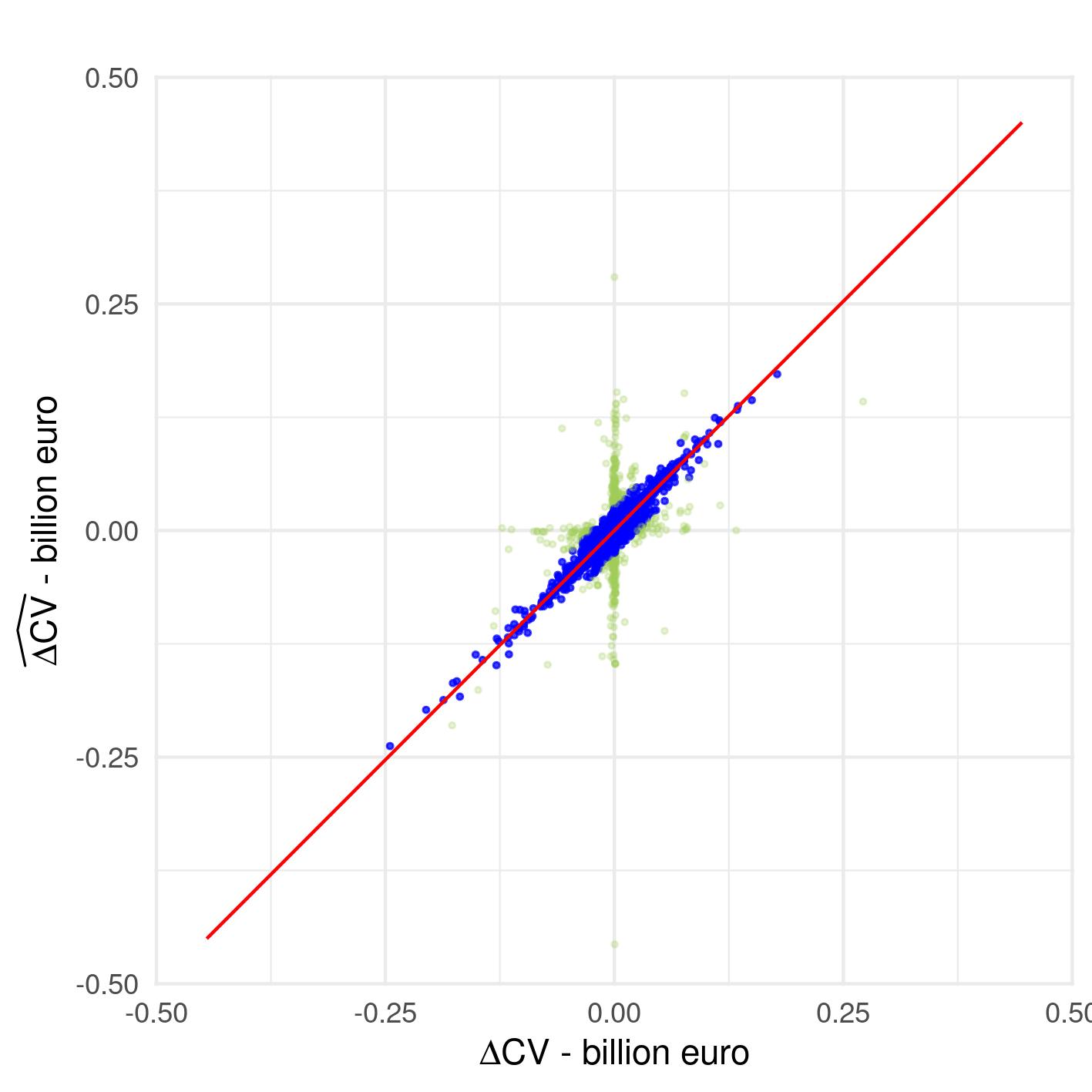} 
	\end{center}
	\vspace{-0.5cm}
	{\footnotesize 
		\begin{spacing}{1}
			Notes: The figure displays the fit between weekly changes in the swap contract values reported in EMIR, $\Delta CV$, and those implied by the pricing model (\ref{eq:swap_value}), $\widehat{\Delta CV}$. Each dot represents a swap-week observation (14,441,405 in total). Green dots denote observations (458 in total) with absolute pricing error greater or equal to \euro 25 million. The regression line, in red, is estimated to have slope of 1.012 with robust method.   
	\end{spacing}}
\end{figure}

Since sensitivity analysis, such as the one at the centre of this study, gives indications about asset price changes, we now discuss the accuracy of the model (\ref{eq:swap_value}) in estimating the change in value of a swap position between two consecutive dates. Let $\Delta CV$ and $\widehat{\Delta CV}$ denote the weekly changes in, respectively, the EMIR reported and model-implied swap contract values. The goodness of the model in fitting the weekly value changes is shown in Figure \ref{fig:DeltaCV_fit_scatter}, where each dot represents a swap-week observation in our dataset. 

\begin{figure}[ht!] 
\caption{Weekly changes in swap contract values and model fitting errors}
\label{fig:error_DeltaCV}
\vspace{-0.2cm}
\begin{center}
	\subfigure[~~~~~~~~~~~~~~~~~~~~~~~~~~~~~~~~~~~~~~~~~~~~~~~~~~~~~~~~~~~~~~~$|\Delta CV|$]{
		
		\includegraphics[width=16cm, height=3.8cm]{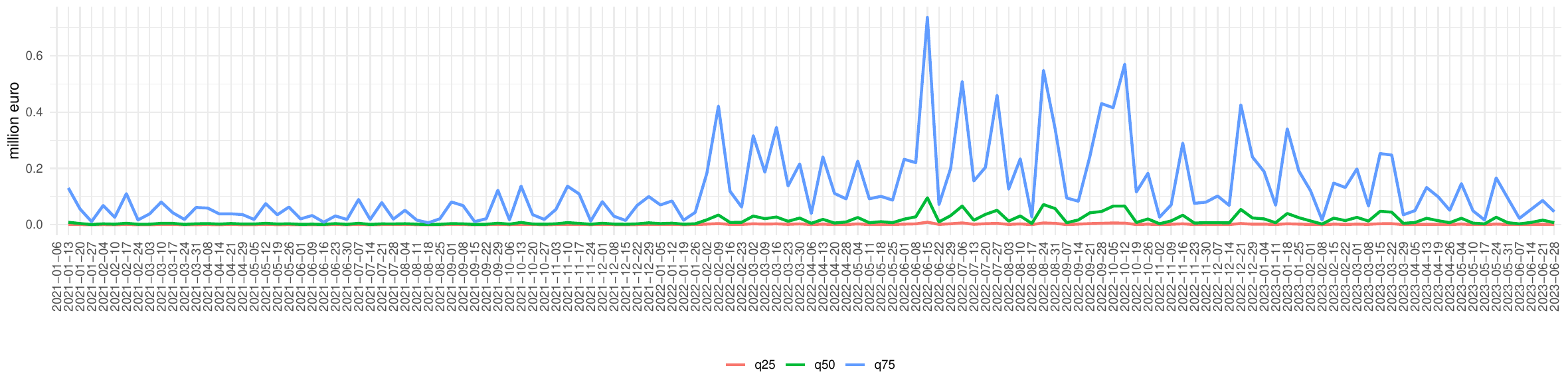}  
	}
	\\ 
	~\\
	\subfigure[~~~~~~~~~~~~~~~~~~~~~~~~~~~~~~~~~~~~~~~~~~~~~~~~~~~~~~~~~~~~$|\Delta CV - \widehat{\Delta CV}|$]{
		\includegraphics[width=16cm, height=3.8cm]{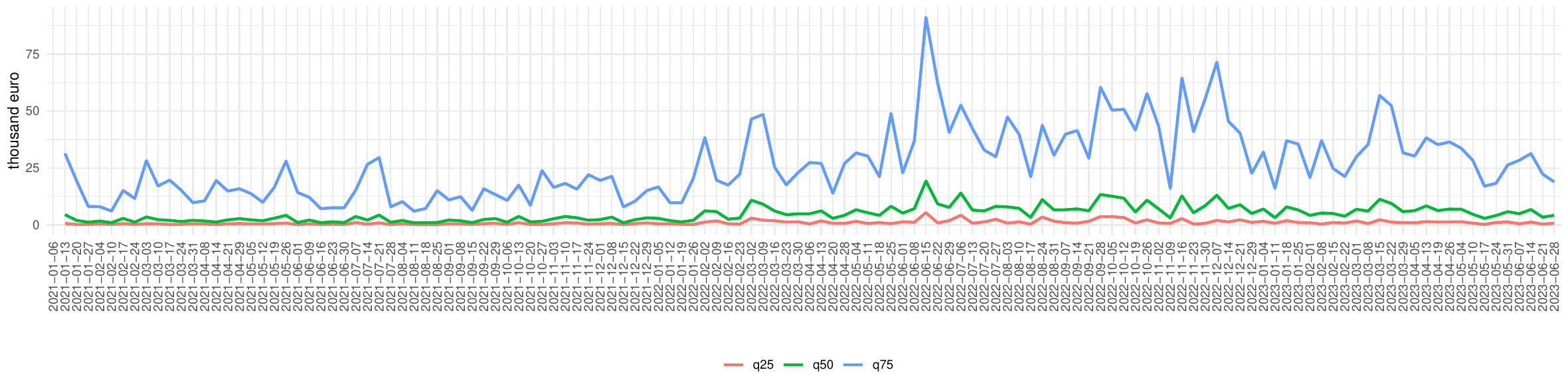} 
	}
\end{center}  
\vspace{-0.5cm}
{\footnotesize 
\begin{spacing}{1}
	Notes: Panel (a) displays the time trend of the 25th, 50th, and 75th percentiles of the distribution of the absolute value of weekly changes in contract values reported in EMIR. Panel (b) displays the time trend of the 25th, 50th, and 75th percentiles of the absolute difference between the weekly changes in the contract values reported in EMIR and those implied by the model (\ref{eq:swap_value}). 
\end{spacing}}	
\end{figure}

\noindent Overall we find that the model fit is improved when the change in value of a swap position, rather than the value itself, is considered. Graphically this means that the vast majority of points in Figure \ref{fig:DeltaCV_fit_scatter} lie on, or are visually indistinguishable from, the regression line, which is estimated to have a slope of 1.012.\footnote{A quick inspection of the outliers reveals that misreporting in EMIR is a contributory factor of the largest deviations observed.} 
As indicated by the less pronounced presence of green dots signalling outliers, the model fitting errors appear smaller than in Figure \ref{fig:CV_fit_scatter}, a fact that is also confirmed by Panel (b) of Figure \ref{fig:error_DeltaCV}, which displays the time trend of the interquartile range of the absolute errors in estimating swap value changes.
Over the full sample period, the median absolute error is \euro 4,330, while the 25th and 75th percentiles amount to \euro 706 and 25,266, respectively. The volatility over time in the interquartile range of the errors seems to track closely that in the interquartile range of the weekly changes in the EMIR swap values, which is shown in Panel (a) of Figure \ref{fig:error_DeltaCV}. 

\clearpage \newpage 

\begin{figure}[h!] 
\caption{Swap portfolio values of the Italian banking system and model fitting errors}
\label{fig:IRS_pricing_fit_system}
\vspace{0.1cm}
\subfigure[Swap portfolio values]{
	\includegraphics[width=7.7cm, height=5.0cm]{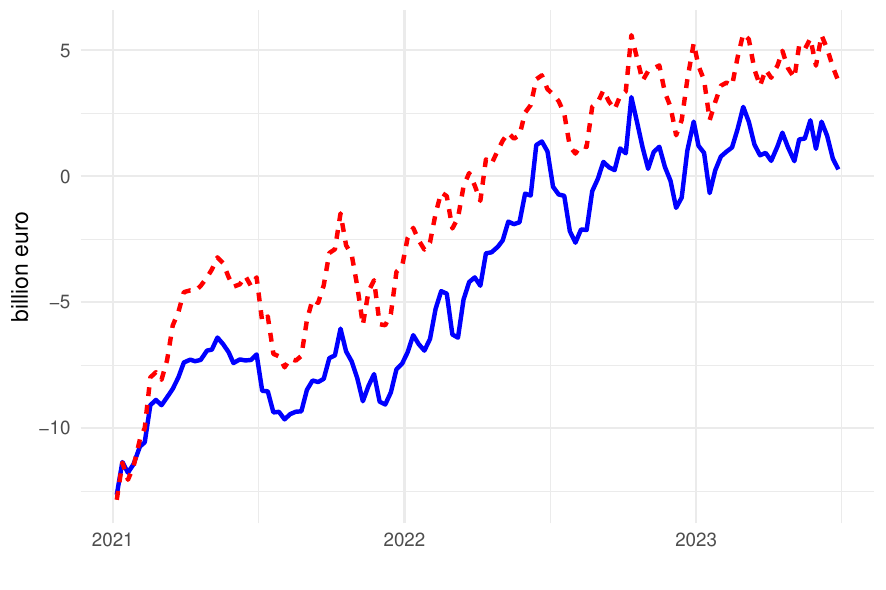}  
}~~ 
\subfigure[Weekly changes in  portfolio values]{
	\includegraphics[width=7.7cm, height=6.0cm]{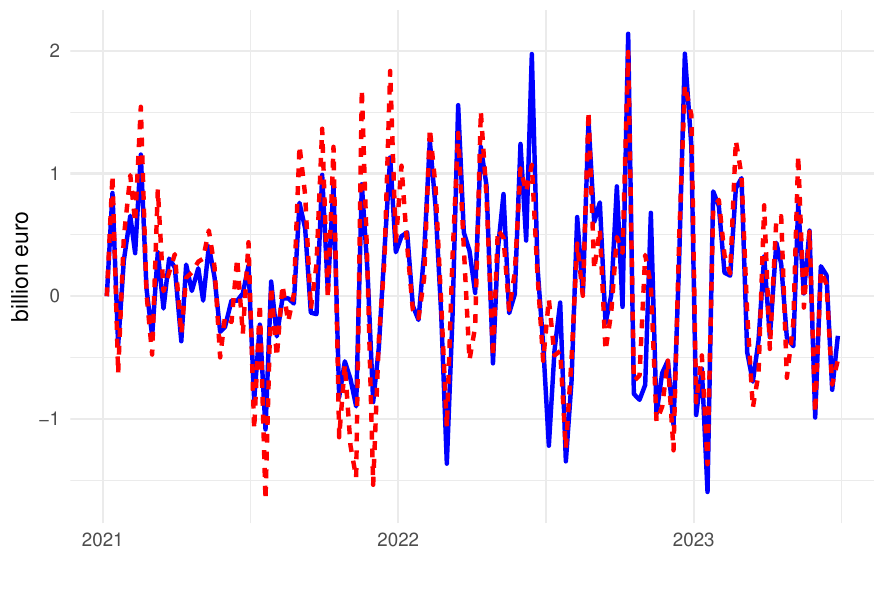} 
}
{\footnotesize 
\begin{spacing}{1}
	Notes: Panel (a) displays the values of the Italian banks' aggregate swap portfolio observed in EMIR (solid blue line) and implied by the model (\ref{eq:swap_value}) (dashed red line). Weekly changes in the aggregate swap portfolio values observed in EMIR and implied by the model are reported in Panel (b). 
\end{spacing}}	
\end{figure}

So far we have analysed the performance of the swap pricing model (\ref{eq:swap_value}) at the level of individual contracts. We now adopt a broader perspective to investigate the model fit on the aggregate swap portfolio value of the Italian banking system, that is the sum of the values of all swap positions of the Italian banks in the sample on a given date. The time trend of the observed and model-implied values of the aggregate swap portfolio are displayed in Panel (a) of Figure \ref{fig:IRS_pricing_fit_system}. One can clearly see that not far from the start of the sample a spread of about \euro 2.5 billion appears between the two series and remains relatively stable throughout the sample. These systematic deviations point to the need for further investigation of the derivatives information collected in EMIR, potentially with the involvement of the reporting entities.    
On the positive side we show, in Panel (b) of Figure \ref{fig:IRS_pricing_fit_system}, that the pricing fit is extremely satisfactory when the weekly changes of the swap portfolio values are considered. The observed and model-implied series track each other quite closely and are often visually indistinguishable.

Taken together, the results of this section suggest that, despite model inaccuracy in pricing a not negligible number of trades, the errors in fitting swap value changes over time are significantly reduced compared to those in fitting the values themselves. This provides compelling evidence that validates the sensitivity analysis discussed below and based on the model-implied changes in swap values following an interest rate shock. 


\subsection{Sensitivity analysis}
\label{sec:sensitivity}

We now employ the tools and data described in the previous sections to study the extent to which banks use swaps to hedge the interest rate risk of their fixed-rate bonds. We do so by computing per-contract valuation changes of a bank's swap and bond portfolios after the interest rate shock, and by comparing the two portfolio impacts to determine whether, in part or in total, they offset each other. Finally, we evaluate the capacity to absorb risk by measuring the portfolio loss (or profit) with respect to bank common equity tier 1 (CET1), to which we simply refer as {\it equity}. 

\begin{figure}[h!] 
\caption{Interest rate shock impact on the Italian banking system} 
\label{fig:sensitivity_system}
\vspace{0.1cm}
\subfigure[Swap and bond sensitivity]{
	\includegraphics[width=7.7cm, height=6.0cm]{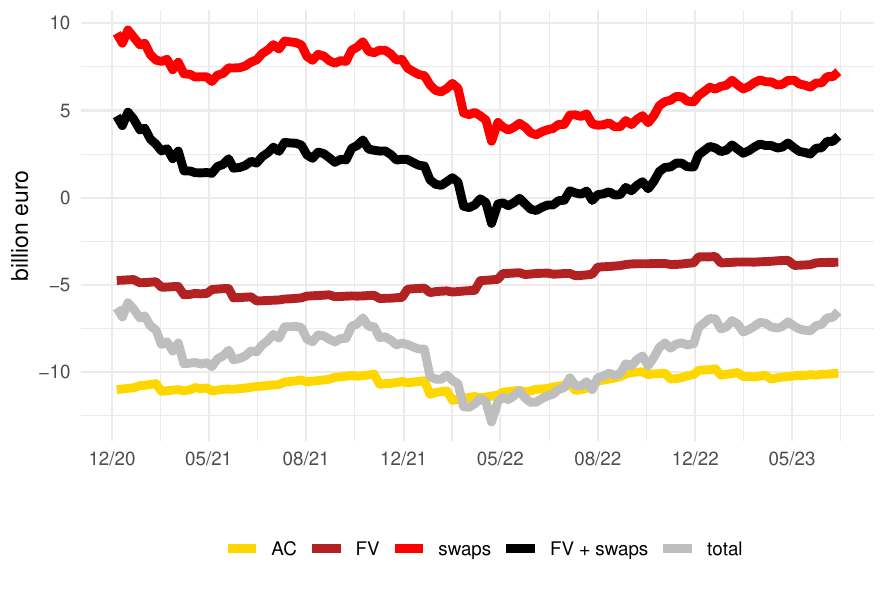}  
}~~ 
\subfigure[Ratio of swap and bond sensitivity to equity]{
	\includegraphics[width=7.7cm, height=6.0cm]{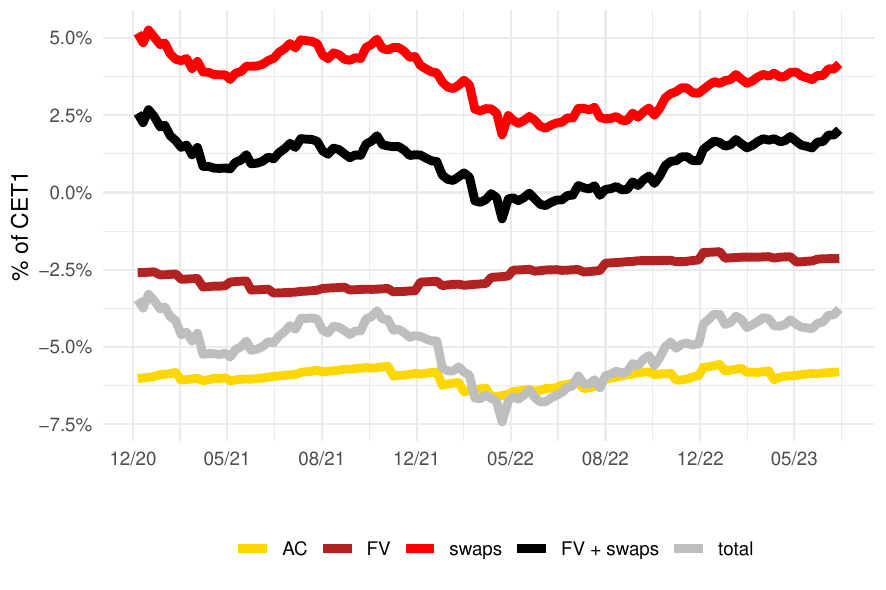} 
}
{\footnotesize 
\begin{spacing}{1}
	Notes: The figure shows the impact of a parallel upward shift of 100 basis points in the yield curve on the aggregate Italian banking system's portfolio of fixed-rate bonds valued at amortised cost (yellow), portfolio of fixed-rate bonds measured at fair value (brown), portfolio of swaps (red), the latter two portfolios (grey), and all portfolios jointly considered (black). The impact is measured in EUR billion in Panel (a), and relative to bank equity in Panel (b). All data are weekly from the first Wednesday of January 2021 to the last of June 2023. 
\end{spacing}}	
\end{figure}

We start by examining the interest rate exposure of the aggregate Italian banking system, for which we display in Figure \ref{fig:sensitivity_system} the time trend of the interest rate shock impact on the portfolio of fixed-rate bonds valued at amortised cost (\textit{AC}), the portfolio of fixed-rate bonds measured at fair value (\textit{FV}), the portfolio of swaps (\textit{swaps}), the two latter portfolios ({\it FV + swaps}), and all portfolios jointly considered (\textit{total}). 

\begin{figure}[!htb]
	\caption{Swap sensitivity by maturity} 
	\label{fig:swap_DV01_maturities}
	\vspace{-0.5cm}
	\begin{center}
		\includegraphics[width=7.7cm, height=6cm]{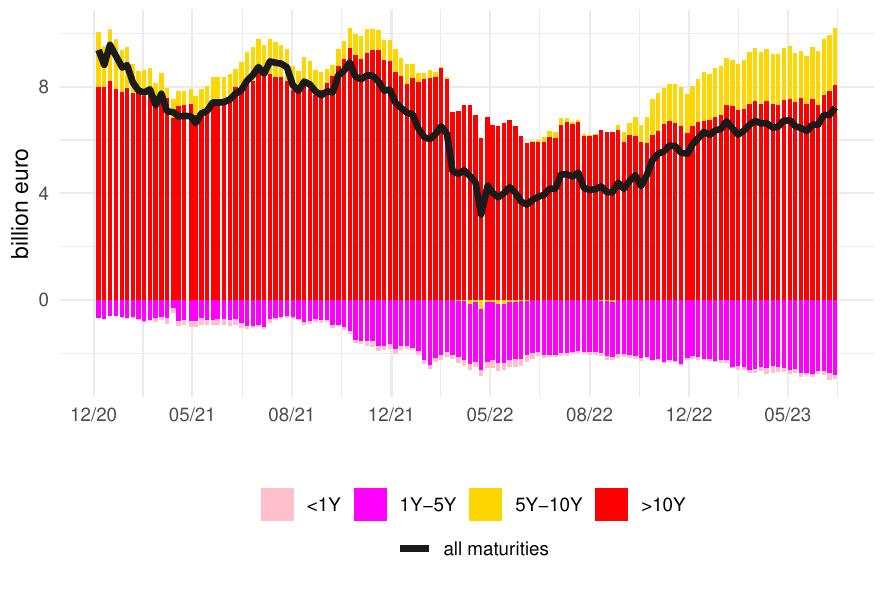} 
	\end{center}
	\vspace{-0.5cm}
	{\footnotesize 
		\begin{spacing}{1}
			Notes: The figure displays the impact of a parallel upward shift of 100 basis points in the yield curve on the aggregate Italian banking system's portfolio of swaps when considered as a whole (solid black line) and when aggregated in four time-to-maturity buckets; values are in EUR billion.
	\end{spacing}}
\end{figure}

\noindent We find that a parallel upward shift of 100 basis points in the yield curve always leads to a profit on the swap portfolio and, on the contrary and not surprisingly, a loss on the bond portfolios (as bond prices move inversely to yields). In particular, it can be seen from Panel (a) that the impact of the positive rate shock hovers around \euro 6.5 billion for swaps, $-$ \euro 4.7 billion for FV bonds, and $-$ \euro 10.8 billion for AC bonds. The finding of a positive impact on the swap portfolio is not trivial if one considers that the aggregate net notional of Figure \ref{fig:IRS_notional} depicts on many occasions the Italian banking system as net short, and we know that the value of a short swap position falls following a rise in rates. The result can however be explained in light of the positive relationship between contract maturity and sensitivity: Italian banks are net long on the longer-dated swaps (over 5 and most importantly 10 years), and when rates rise these positions appreciate in value with profits that subdue the losses on the shorter-dated swaps (between 1 and 5 years), on which banks are net short with larger yet less sensitive notional amounts. This is shown graphically in Figure \ref{fig:swap_DV01_maturities}, from which we see that the sensitivity of the banks' swap portfolio is almost completely determined by longer-dated contracts, proving that (gross and net) notional exposures are not adequate measures of risk.

Another interesting finding from Figure \ref{fig:sensitivity_system} is that the aggregate gains on swaps fully offset the losses on bonds carried at fair value, indicating that banks may use swaps to hedge the value of these assets against changes in rates.
However, this does not seem to hold true for losses on AC bonds, and, as a consequence, the total impact of the rate shock on the Italian banking system is negative, ranging from $-$ \euro 7 to $-$ \euro 12.5 billion. Intuitively, if one considers that AC bonds are valued at amortised cost because the intent is to hold these securities until maturity, it is natural to expect that the interest rate risk hedge is in place first and foremost for the FV bonds. The important role of swaps in hedging banks' interest rate risk exposures can best be appreciated by analysis of the rate shock impact on banks' equity. From Panel (b) of Figure \ref{fig:sensitivity_system} (grey line), we see that, on average, the total impact of a 100 basis point upward shift in the yield curve is reduced from a loss of approximately 8.6\% of CET1 to a loss of 5\% after accounting for banks' swap positions.
It is important to stress that these reductions of bank equity would only materialise in the unlikely scenario in which financial intermediaries are forced to sell their portfolio of AC debt securities before maturity. Therefore in practice it is extremely likely that, as interest rates change, bank equity is only affected by variations in the market values of swaps and FV bonds, and the results in Figure \ref{fig:sensitivity_system} (black line) suggest that on average and in aggregate the effect on swaps dominates, with a favourable outcome for bank equity when rates rise (the loss turns into a gain of 1.0\% of CET1 when AC bonds are not considered). Based on this evidence, we can state that, on aggregate, swaps are of economic importance for Italian banks to hedge and absorb risk from their fixed-rate bonds. This contrasts with some recent results reported for US banks by \cite{McPhail2023_US_swaps}, who conclude that swap positions are not economically meaningful in hedging the interest rate risk of bank assets. 


We now explore the heterogeneity of our findings across institutions.  
Table \ref{tab:summary_stas} presents summary statistics on swap and bond portfolios computed at the level of individual banks; for completeness we also report statistics at the aggregate sector level (column 1). The latter has been at the centre of our considerations so far and therefore we do not comment any further on this. Now turning to the bank-level statistics in Panel A, which are obtained for the full sample of 54 banks considered in this work, we observe that the average bank has an economically meaningful interest rate risk from swaps. Following a 100-basis-point increase in rates, the swaps of the average bank appreciate in value by \euro 100 million, or about 5.6\% of bank equity, which partly offsets the value losses of  \euro 200 and \euro 100 million on AC and FV bonds, respectively. We find considerable dispersion across banks: the rate shock impact on swaps varies from $-$ \euro 200 million at the 5-th percentile to \euro 700 million at the 95-th percentile, and the ratio of such impact to bank equity varies from -6\% to 28.5\%. The distribution of the ratio of the rate impact on swaps to equity has a positive median (1.3\%), implying that the direction in which the majority of Italian banks trade swaps, that is by benefiting from an increase in market interest rates, is consistent with hedging their business risk of borrowing short and lending long (maturity mismatch). This is also in line with evidence on swap trades provided for the European banking sector by \cite{hoffmann2019bears} and \cite{ECB_FSR_Nov22}, and for UK banks by \cite{khetan2023market}.
As for the sensitivity of fixed-rate bonds, not surprisingly we find that the impact is always non-positive (as bond prices move inversely to yields): the rate shock impact on AC bonds varies from $-$ \euro 1.1 billion at the 5-th percentile to nil at the 95-th percentile, while the impact on FV bonds varies from $-$ \euro 300 million to nil.
By combining the sensitivity of swaps and bonds together, we find that the ratio of the rate impact on all portfolios to bank equity varies from -22.1\% at the 5-th percentile to 11.9\% at the 95-th percentile. When we only consider swap and FV bond portfolios, the impact relative to equity varies from -8.3\% to 26.1\%. Examining Panels B and C of Table \ref{tab:summary_stas}, which report statistics for the restricted sample of, respectively, significant institutions (SI) and less significant institutions (LSI), two 
important remarks can be made. 

\begin{table}[!ht]
\caption{Summary statistics}
\label{tab:summary_stas}
\fontsize{9}{11} \selectfont
\renewcommand{\tabcolsep}{5.2pt} 
\centering 
\begin{tabular}{lSSSSSS} \hline 
	\vspace{-0.15cm}\\
	 & {Aggregate} & {Mean} & {St. Dev.}  & {Median} & {p5} & {p95} \vspace{+0.12cm}\\
	\hline
	\multicolumn{7}{l}{\rule[-2.2mm]{0mm}{6.5mm}{\emph{\textbf{Panel A -- all banks}}}} \\	
	\hline
	\vspace{-0.15cm}\\			
	{Swap Notional} & 5346.6 & 97.2 & 455.9 & 0.2 & 0.0 & 134.0 \\
	{Swap Net Long Notional} & -33.6 & -0.6 & 9.1 & 0.0 & -4.9 & 6.3 \\
	{Swap Market Value} & -3.8 & -0.1 & 0.6 & 0.0 & -0.6 & 0.4 \\
	{AC + FV Bond Market Value}  & 319.2 & 5.7 & 11.8 & 0.9 & 0.1 & 29.2 \\
	{AC Bond Market Value}  & 220.8 & 3.9 & 7.9 & 0.6 & 0.0 & 21.4 \\
	{FV Bond Market Value}  & 98.7 & 1.9 & 4.6 & 0.4 & 0.0 & 7.4 \\
	{Swap DV01} & 6.5 & 0.1 & 1.0 & 0.0 & -0.2 & 0.7 \\
	{AC Bond DV01}  & -10.8 & -0.2 & 0.4 & 0.0 & -1.1 & 0.0 \\
	{FV Bond DV01}  & -4.7 & -0.1 & 0.3 & 0.0 & -0.3 & 0.0 \\
	{Ratio: Swap DV01 to Equity} & 3.6 &  5.6  & 13.9  & 1.3  & -6.0  & 28.5  \\
	{Ratio: Swap + FV Bond DV01 to Equity} & 1.0  & 3.2 & 14.0 & -0.1 & -8.3 & 26.1 \\
	{Ratio: Total DV01 to Equity}  & -5.0  & -5.3 & 14.3 & -4.8 & -22.1 & 11.9 \\
	\vspace{-0.1cm}\\  
	\hline
	\\[-0.3cm]	
	\multicolumn{7}{l}{\rule[-2.2mm]{0mm}{6.5mm}{\emph{\textbf{Panel B -- SI banks }}}} \\
	\hline
	\vspace{-0.15cm}\\			
	{Swap Notional} & 5276.2 & 405.9 & 894.4 & 17.5 & 0.3 & 2412.8 \\
	{Swap Net Long Notional} & -27.4 & -2.1 & 18.9 & 3.0 & -27.4 & 8.1 \\
	{Swap Market Value} & -1.6 & -0.1 & 1.0 & 0.0 & -1.6 & 0.7 \\
	{AC + FV Bond Market Value}  & 264.7 & 20.4 & 18.0 & 18.3 & 2.4 & 53.9 \\
	{AC Bond Market Value}  & 182.9 & 14.0 & 11.5 & 13.9 & 1.3 & 32.0 \\
	{FV Bond Market Value}  & 81.9 & 6.3 & 7.8 & 3.6 & 0.4 & 21.5 \\
	{Swap DV01} & 6.6 & 0.5 & 2.0 & 0.3 & -1.4 & 3.2 \\
	{AC Bond DV01}  & -9.3 & -0.7 & 0.7 & -0.7 & -1.7 & -0.0 \\
	{FV Bond DV01}  & -4.2 & -0.3 & 0.6 & -0.1 & -1.3 & -0.0 \\
	{Ratio: Swap DV01 to Equity} & 4.3 &  8.0  & 11.5  & 5.7  & -3.6  & 27.1  \\
	{Ratio: Swap + FV Bond DV01 to Equity} & 1.5  & 5.7 & 11.8 & 4.9 & -4.9 & 26.1 \\
	{Ratio: Total DV01 to Equity}  & -4.5  & -8.7 & 20.8 & -4.0 & -36.9 & 7.1 \\
	\vspace{-0.1cm}\\  
	\hline
	\\[-0.3cm]	
	\multicolumn{7}{l}{\rule[-2.2mm]{0mm}{6.5mm}{\emph{\textbf{Panel C -- LSI banks}}}} \\
	\hline
	\vspace{-0.15cm}\\			
	{Swap Notional} & 21.2 & 0.6 & 1.3 & 0.1 & 0.0 & 2.8 \\
	{Swap Net Long Notional} & 12.3 & 0.4 & 1.0 & 0.0 & -0.1 & 2.0 \\
	{Swap Market Value} & -0.7 & -0.0 & 0.2 & 0.0 & -0.0 & 0.1 \\
	{AC + FV Bond Market Value}  & 39.7 & 1.1 & 1.2 & 0.6 & 0.1 & 3.0 \\
	{AC Bond Market Value}  & 26.0 & 0.7 & 0.9 & 0.3 & 0.0 & 2.0 \\
	{FV Bond Market Value}  & 13.8 & 0.4 & 0.5 & 0.3 & 0.0 & 1.2 \\
	{Swap DV01} & 1.0 & 0.0 & 0.1 & 0.0 & -0.0 & 0.2 \\
	{AC Bond DV01}  & -1.1 & -0.0 & 0.0 & 0.0 & -0.1 & -0.0 \\
	{FV Bond DV01}  & -0.4 & -0.0 & 0.0 & 0.0 & -0.0 & -0.0 \\
	{Ratio: Swap DV01 to Equity} & 7.8 &  6.0  & 14.4  & 0.9  & -0.5  & 27.6  \\
	{Ratio: Swap + FV Bond DV01 to Equity} & 5.0  & 3.4 & 14.7 & -0.1 & -6.7 & 23.2 \\
	{Ratio: Total DV01 to Equity}  & -5.0  & -3.6 & 11.6 & -4.8 & -18.7 & 16.6 \\
	\vspace{-0.15cm}\\
	\hline
	\vspace{-2.5mm}
\end{tabular}
\begin{minipage}{15.8cm}
	Notes: This table provides summary statistics of banks' swap and bond portfolios. All values are in EUR billion, except for the ratios that are in percentage. Aggregate amounts are computed by first summing across banks on each date and then averaging across dates, whilst the remaining statistics are computed by pulling banks and dates together. Panel A reports statistics for the full sample of 54 banks, which includes 4 subsidiaries of non-Italian banks. Panels B and C report statistics for, respectively, significant (SI) and less significant (LSI) Italian institutions. Swap indicates the portfolio of euro interest rate swaps (IRS, OIS, and FRA), AC denotes the portfolio of euro-denominated fixed-rate bonds valued at amortised cost, and FV denotes the portfolio of euro-denominated fixed-rate bonds measured at fair value. Equity is CET1 equity. DV01 is the impact of a 100-basis-point parallel upward shift in the yield curve. Data are weekly and sampled each Wednesday from January 2021 to June 2023. 
\end{minipage}
\end{table}

\clearpage \newpage

\noindent First, variation exists also across SI banks and, to a lesser extent, LSI banks, with some swap positions playing an offsetting role and some exacerbating bond market exposures to interest rate risk. However, it is important to note that the overall risk exposure of a given bank in derivatives may be driven by contracts other than swaps, such as options and futures on interest rates and bonds; the inclusion of these and other contracts is high in our research agenda. Second, the use of derivatives by LSI banks seems quite limited with their aggregate swap sensitivity estimated to be on average \euro 1 billion compared to the \euro 6.6 billion aggregate swap sensitivity of SI banks.   


From these results, and based on supervisory reports (FINREP), we are able to assert that some of the banks implementing micro fair value hedge accounting strategies can rely on swaps to hedge the interest risk of their entire fixed-rate bond portfolio. In fact, for those banks the profits on swaps following the increase in rates absorb not only the losses on FV bonds but also those on AC bonds.


\section{Conclusions}
\label{sec:Conclusions}

The objective of this paper is twofold. First, our analysis of interest rate risk in Italian banks underscores the ever-present challenge of managing financial risks in a dynamic economic landscape. We develop a tool to determine the fair value of bond and derivative portfolios under different scenarios and analyse their interest rate exposures from a risk management perspective. Our framework relies on granular information on these assets, including specific contract characteristics. 
Our empirical findings show that, for Italian banks, an upward shift of 100 basis points in the yield curve causes material losses on bonds, which are partially offset by the positions on euro interest rate swaps. These losses decrease significantly when bonds valued at amortised cost are disregarded: in that case, we observe a favourable outcome for bank equity as the profits on swaps are, on average, of greater size than the losses on bonds measured at fair value. 

Second, the investigation of data quality issues associated with the ``value of contract'' field reported in EMIR has revealed a critical area for improvement. Our contract-by-contract full revaluation approach allow us to conclude that, despite model inaccuracy in pricing a not negligible number of trades, the errors in fitting swap value changes over time are significantly reduced compared to those in fitting the values themselves. This work represents a pioneering effort as it is the first extensive data quality analysis conducted on this field. This groundbreaking initiative establishes a foundational benchmark for rigorously evaluating and improving data accuracy in this pivotal aspect of EMIR reporting, marking a significant advancement on this research direction. Accurate and reliable data is the core of effective risk assessment. Resolving these data quality issues is not merely a matter of compliance but is crucial for a sound regulatory oversight and financial stability monitoring.

\clearpage \newpage 


\clearpage \newpage 
\setcounter{section}{0} 

\begin{center}
	\LARGE \textbf{Appendix}
\end{center}

\renewcommand{\thesection}{A} \setcounter{equation}{0} 

\renewcommand\thefigure{\thesection.\arabic{figure}} 
\renewcommand\thetable{\thesection.\arabic{table}}       
\setcounter{figure}{0} \setcounter{table}{0}

\section{Bootstrapping the yield curve}
\label{sec:bootstrap_rates} 

We review the procedure for bootstrapping the riskless spot (i.e.~zero-coupon) curve that provides discount rates of swap and bond pricing and is the object of the shock in our sensitivity analysis.\footnote{In the case of derivatives, including swaps, discount rates and risk-free rates are the same, whereas the discount rates of bonds have an extra component (spread) that reflects the credit risk of the issuer.}
Following the standard multi-curve method outlined by \cite{Hull2015} and \cite{Smith2013}, among others, we bootstrap spot rates from \euro STR-referencing OIS rates, and from EURIBOR swap rates. To this end, we obtain from Refinitiv daily rate observations for \euro STR OIS with maturity up to 30 years, and for 6-month EURIBOR swaps with maturity up to 50 years. These are ``par'' (at-market) swap rates representing the fixed-rate paid on bonds valued at par (i.e.~price of 100).\footnote{The par swap rates are also the rates that make the market value of the swap contracts equal zero.} We linearly interpolate the swap rates to have data for maturities evenly spaced by 3-month intervals as we assume quarterly settlements of the swap contracts and, accordingly, four payments a year on the par bonds. OIS swaps of up to one-year's maturity have only a single payment at the contract end date, therefore, we use their rates as the spot rates of the corresponding maturity. 

To infer the rest of the spot curve that is consistent with the sequence of OIS par swaps we rely on the following bootstrapping technique.  Let $z^{(T)}$ be the annualised spot interest rate observed today for maturity $T$ (in months), $s^{(T)}$ be the annualised par swap rate for maturity $T$, and $P^{(T)}$ be the $T$-maturity par bond's price, which is equal to its notional amount, $N=100$. Under the assumption of quarterly settlement, each coupon payment of the par bond amounts to $C = s^{(T)} N / 4$. Using the spot rates already available for all maturities before $T$, the next-in-line spot rate $z^{(T)}$ is the solution to the following bond pricing equation
\begin{equation}
	\label{eq:bond_price}
	P^{(T)} =  \sum_{n \in D} \frac{C}{(1 + z^{(n)} \frac{n}{12})}  + \frac{C + N}{(1 + z^{(T)} \frac{T}{12})} ~ ,
\end{equation} 
where $D=\{3,6,...,T-3\}$. Equation (\ref{eq:bond_price}) implies that the spot rate gets computed as
\begin{equation}
	\label{eq:spot_rate}
	z^{(T)} = \frac{12}{T}  \bigg[ \frac{C + N}{P^{(T)} - \sum_{n \in D} \frac{C}{(1 + z^{(n)} \frac{n}{12})} } - 1 \bigg] ~ .
\end{equation} 

\noindent We repeat the calculation sequentially until the longest maturity of the par swap rates, and then we linearly interpolate the spot rates to have data for maturities evenly spaced by 1-month intervals.\footnote{When discounting the future cash flows of the swap contracts sampled from EMIR, despite using an actual/365 convention, we will round to the nearest month the maturity of the relevant spot rates.}
We use the above procedure to bootstrap one spot curve from OIS rates and one from EURIBOR swap rates. 
The former spot curve provides discount factors for maturities up to 30 years, which is the longest maturity observed in the \euro STR OIS market, and determines the forward rates affecting the cash flows of the OIS contracts sampled from EMIR.\footnote{EONIA was discontinued on 3 January 2022 and therefore we use \euro STR rates to value also the EONIA-referencing OIS contracts that continue to exist until their expiration.}
The latter spot curve provides discount factors for maturities longer than 30 years and determines the forward rates affecting the cash flows of the FRA and IRS contracts sampled from EMIR. 
We compute EURIBOR and \euro STR forward rates, which represent the expected future rates on the floating leg of the swap contracts considered in our work, starting from the corresponding spot curve and assuming the absence of arbitrage.
Letting $f^{(T_{i-1},T_{i})}$ be the forward rate between times $T_{i-1}$ and $T_{i}$ (in months) -- i.e.~the interest rate expected today on a zero-coupon investment starting at time $T_{i-1}$ and ending at $T_{i}$ -- the assumption of no-arbitrage implies that the following equality holds
\begin{equation}
	\label{eq:forward_rate}
	f^{(T_{i-1},T_{i})} = \frac{12}{T_{i}-T_{i-1}}  \bigg( \frac{ 1 + z^{(T_{i})} \frac{T_{i}}{12} } { 1 + z^{(T_{i-1})} \frac{T_{i-1}}{12} } - 1 \bigg) ~ .
\end{equation} 

\clearpage \newpage

\renewcommand{\thesection}{B} \setcounter{equation}{0} 

\renewcommand\thefigure{\thesection.\arabic{figure}} 
\renewcommand\thetable{\thesection.\arabic{table}}       
\setcounter{figure}{0} \setcounter{table}{0}

\section{Pricing formulae}
\label{sec:pricing_formulae} 

\subsection{Swap pricing}

Besides the fixed and floating rates discussed in Section \ref{sec:EMIR_Data}, the other inputs of our swap pricing model (\ref{eq:swap_value}) consist of the contractual features that are reported in the following EMIR fields: ``valuation timestamp'', ``counterparty side'', ``notional'', ``maturity date'', ``fixed-rate payment frequency'', and ``floating rate payment frequency''. We also pull the ``value of contract'' field which we use to assess the fit of our pricing functions.
As a first step in determining the two bond prices in  (\ref{eq:swap_value}), we define the payment schedule of the fixed and floating legs by using the payment frequency expressed in months and working backwards from maturity date $T$ to valuation date $t_0$. For each payment schedule, we compute, both in days and in months, the time frame between $t_0$ and each of the future payment dates.\footnote{In order to have integer numbers for the number of months in the time frame, we shift forward both the valuation and the next payment dates to month-end dates and we count the months between the shifted dates. If the first next payment date is within a month, we set the time frame to 1 month.} We will use the time frames expressed in days to perform the actual/365 day-count convention for rates when discounting, and those expressed in months to select the maturity of the relevant spot rate from those interpolated at 1-month intervals. We cap all time frames between payment dates and $t_0$ to 50 years as this corresponds to the longest maturity of the bootstrapped spot rates. 

To calculate the value of a swap, we implement the following pricing formulae. Let us consider an interest rate swap referencing EURIBOR with maturity $T$ and notional amount $N$. The fixed leg, which pays the annualised swap rate $s^{(T)}$, makes $q^{fix}$ payment(s) per year, for a total of $I$ payments between $t_0$ and $T$. Let $n_i$ and $d_i$ be the time frames in months and in days, respectively, between $t_0$ and the $i$-th fixed-rate payment date, with $i=1,2,...,I$. The floating leg, which pays the annualised $k$-month EURIBOR rate, with $k=1,3,6,12$, makes $q^{fl}$ payment(s) per year, for a total of $J$ payments between $t_0$ and $T$. Let $n_j$ and $d_j$ be the time frames in months and in days, respectively, between $t_0$ and the $j$-th floating-rate payment date, with $j=1,2,...,J$. We assume that at each payment date $j$ the floating rate bond pays the reference rate that prevailed on the market at the previous payment date, $j-1$. This means that the first floating rate payment is already known at valuation date and is based on the EURIBOR spot rate observed $n_1 - k$ months before $t_0$. Using the spot rates $z^{(\cdot)}$ defined in Section \ref{sec:Methodology}, the bond price of the fixed leg gets computed as
\begin{equation}
	\label{eq:fix_bond_price}
	B^{fix} =  \sum_{i=1}^{I} \frac{C^{fix}}{(1 + z^{(n_i)} \frac{d_i}{365})}  + \frac{N}{(1 + z^{(n_I)} \frac{d_I}{365})} ~ ,
\end{equation} 
where $C^{fix}=s^{(T)}N/q^{fix}$. At the same time, while relying on both spot and forward rates, the bond price of the floating leg gets computed as
\begin{equation}
	\label{eq:fl_bond_price}
	B^{fl} =  \sum_{j=1}^{J} \frac{C^{fl}_j}{(1 + z^{(n_j)} \frac{d_j}{365})}  + \frac{N}{(1 + z^{(n_J)} \frac{d_J}{365})} ~ ,
\end{equation}
where for $j=1$ we use $C^{fl}_1=z^{(k)}_{t_{0}^{-}} N/q^{fl}$, with $z^{(k)}_{t_{0}^{-}}$ denoting the $k$-month EURIBOR spot rate observed $n_1 - k$ months before $t_0$, and for $j>1$ we use $C^{fl}_j=f^{(n_{j-1},n_{j})}N/q^{fl}$, with $f^{(n_{j-1},n_{j})}$ denoting the $k$-month EURIBOR forward rate between times $n_{j-1}$ and $n_{j}$. By combining the two bond prices as per equation (\ref{eq:swap_value}), we obtain the model-implied values of the IRS contracts in this work. We now review the adjustments we make to model (\ref{eq:swap_value}) for the other contracts considered. 

In the case of FRA, which are single-period contracts, we value a short (i.e.~receive-fixed) position as the difference between the agreed-upon fixed-rate and the forward rate computed at valuation date. The rate difference is first multiplied by the notional and the length of time (in years) between the FRA effective date, which corresponds to the beginning of the forward-starting loan, and maturity date, which corresponds to the end of it, and then discounted back to valuation date. 

In the case of OIS, the first adjustment that we make accounts for the fact that the floating leg pays the reference rate compounded daily over a set time period. This means that, whenever valuation date $t_0$ is past the effective date of the contract, the first floating leg payment is given in part by the compounded \euro STR rate that has prevailed in the market from the previous payment date to $t_0$, and in part by the \euro STR spot rate applicable to an investment that begins at $t_0$ and terminates at the next payment date.
The second adjustment accounts for the fact that OIS payments are annual (i.e.~at most one payment per year). This implies that, for spot-starting contracts, we compute the annual payments using 100\% of the fixed and floating rates if maturity date is more than one year away from the contract effective date, otherwise we only use a fraction that is given by the number of months between the two dates, divided by 12. For forward-starting OIS contracts we always use 100\% of rates to compute each of the annual payments.

The valuation approach described in this section delivers the model-implied swap values that we first compare against the EMIR prices to assess the goodness of fit and then we use to compute the banks' swap sensitivities.

\subsection{Bond pricing}

To evaluate the impact of the yield curve shift on the euro-denominated fixed-rate bonds, we consider a partial revaluation approach based on a second-order approximation (deltagamma) relying on modified duration and convexity directly provided by Refinitiv.
Since our sensitivity analysis is focused on euro denominated fixed-rate bonds, among the various characteristics available in the Security Database, we consider only the coupon type and the currency to define the risk factors needed for the selection of assets on which to conduct the empirical analysis. The bonds are assigned to maturity buckets depending on the residual maturity (below 1Y, 1Y-3Y, 3Y-5Y, 5Y-7Y, 7Y-10Y, 10Y-20Y, above 20Y), more granular than in Figure \ref{fig:Portfolios}, and split by security type (i.e.~ordinary bonds, securitizations and covered bonds). This step is needed to defined duration and convexity data for bonds for which this information is not available in Refinitiv.

The profit-and-loss (PnL) of a bond $j$ with fair value $FV^j_t$ at time $t$ under the scenario in which we shift the interest rate curve by $s$ can be computed by taking into account the following equality
\begin{equation}\label{eq:PnL_bonds}
	PnL(t,j) = FV^j_t {r_{ir}}(t,j),
\end{equation} 
where $r$ represents the return corresponding to the interest rate risk. More in details, we have
\begin{equation*}\label{eq:return_ir}
	{r_{ir}^i}(t,j) = - D_t^j s + \frac{1}{2}C_t^j s^2,
\end{equation*} 
where $D_t^j$ and $C_t^j$ are the modified duration and the convexity of the security $j$ at a given date $t$, and $s$ the shift of the interest rate curve. It appears clear that the returns related to interest rates depend on the characteristic of each single security (i.e. duration and convexity), either than the selected shift. 

The overall PnL is a function of the shift, bank exposures and properly selected modified durations and convexities. We extracted from Refinitiv these weekly data (i.e. the fields DM and CX) in the period between December 31, 2020 and June 30, 2023.\footnote{As already observed, since security holdings portfolios are reported only at the end of each month, we assume that the bonds portfolios remains constant during the month while duration and convexity are updated weekly.} For all debt securities for which modified duration and convexity are not available at a given point in time, we consider weighted average durations and convexities. These weighted averages are evaluated  across all intermediaries at each trading date for each maturity bucket and  debt security type (i.e. ordinary bonds, securitizations and covered bonds). It should be highlighted that Refinitiv reports duration and convexity for a large number of bonds. We deal with around 15,000 bonds. Duration and convexity data are available for slightly more than 12,000 bonds, representing 84\% of the total bond fair value. This percentage does not vary significantly during the analysed period. 

\end{document}